%% file: main.tex
\documentclass[letterpaper,twocolumn,10pt]{article}
\usepackage{usenix2020}
\usepackage{url}
\usepackage{multirow}
\usepackage{adjustbox}
\usepackage{listings}
\usepackage{xspace}
\usepackage{amsmath}
\usepackage{booktabs}
\usepackage{comment}
\usepackage{enumitem}
\usepackage{subcaption}
\usepackage{algorithmicx}
\usepackage{algorithm}
\usepackage[noend]{algpseudocode}
\usepackage[normalem]{ulem}
\usepackage{amsthm}

\usepackage{pifont}
\usepackage[textsize=tiny,textwidth=0.6in]{todonotes}

\input{macro}
\begin{document}
\pagestyle{empty}

\title{\tool: Cost-Efficient Multi-LLM Serving via GPU Memory Ballooning}
\author
{
Shan Yu$^{1}$,\enspace
Yifan Qiao$^{2}$,\enspace
Mingyuan Ma$^{3}$,\enspace
Yangmin Li$^{4}$,\enspace
Shuo Yang$^{2}$, \enspace
Xinyuan Tong$^{5}$, \enspace
Yang Wang$^{6}$,\enspace\\
Zhiqiang Xie$^{7}$,\enspace
Yuwei An$^{4}$, \enspace
Shiyi Cao$^{2}$,\enspace
Ke Bao$^{8}$,\enspace
Deepak Vij$^{9}$, \enspace
Xiaoning Ding$^{9}$, \enspace
Yichen Wang$^{9}$, \enspace\\
Qingda Lu$^{10}$, \enspace
Zhong Wang$^{11}$, \enspace
Gao Gao$^{12}$, \enspace
Harry Xu$^{1}$\textsuperscript{*},\enspace
Junyi Shu$^{1}$\textsuperscript{*}, \enspace
Jiarong Xing$^{13}$\textsuperscript{*}, \enspace
Ying Sheng$^{1}$\textsuperscript{*}\\[0.5ex]
\vspace{1mm}
\normalsize
$^1$UCLA \quad
$^2$UC Berkeley \quad
$^3$Harvard University \quad
$^4$CMU \quad
$^5$University of Edinburgh \quad
$^6$Intel  \quad
$^7$Stanford University \\
\normalsize
$^8$LMSYS \quad
$^9$ByteDance \quad
$^{10}$Alibaba Cloud \quad
$^{11}$Tsinghua University\quad
$^{12}$Novita AI\quad
$^{13}$Rice University
}

\maketitle
\blfootnote{* Corresponding authors.}

\input{abstract_v1}
\input{intro_v3}
\input{background}

\input{motivation_v2}
\input{overview}

\input{design-v2}

\input{scheduler-v3}
\input{eval-v1}

\input{rw-v2}

\input{conclusion}

\normalsize
\bibliographystyle{abbrv}
\bibliography{paper}
\clearpage
\appendix
\input{appendix}

\end{document}

%% file: macro.tex
\microtypecontext{spacing=nonfrench}

\newcommand{\tool}[0]{Prism\xspace}
\newcommand{\muxp}[0]{MuxServe++\xspace}

\definecolor{javared}{rgb}{0.6,0,0}
\definecolor{javagreen}{rgb}{0.25,0.5,0.35}
\definecolor{javapurple}{rgb}{0.5,0,0.35}
\definecolor{javadocblue}{rgb}{0.25,0.35,0.75}

\lstdefinestyle{mystyle}{
      language=C++,
        basicstyle=\tiny\ttfamily,
        keywordstyle=\color{javapurple}\bfseries,
        stringstyle=\color{javared},
        commentstyle=\color{javadocblue},
        morecomment=[s][\color{javadocblue}]{/**}{*/},
        numbers=left,
        breaklines=true,
        numberstyle=\tiny\color{black},
        stepnumber=1,
        numbersep=5pt,
        tabsize=2,
        showspaces=false,
        showstringspaces=false,
        morekeywords={foreach, uint64_t, uint16_t},
        classoffset=0,
        xleftmargin=1.8em,
        escapeinside={(*@}{@*)},
        moredelim=*[is][\color{red}]{[[[}{]]]},
        captionpos=b,
}
\newcommand{\naive}[0]{na\"{i}ve\xspace}
\newcommand{\naively}[0]{na\"{i}vely\xspace}
\newcommand{\codeIn}[1]{{\small\texttt{#1}}}

\newcommand{\MyPara}[1]{\vspace{.1em}\noindent\textbf{#1}~}
\newcommand{\code}[1]{{\tt #1}}

\newcommand\blfootnote[1]{%
  \begingroup
  \renewcommand\thefootnote{}\footnote{#1}%
  \addtocounter{footnote}{-1}%
  \endgroup
}
\newcommand{\mysection}[1]{\vspace{-1em}\section{#1}}
\newcommand{\mysubsection}[1]{\vspace{-1em}\subsection{#1}}

\newcommand{\red}[1] {{\textcolor{red}{#1}}}

\newcommand{\kvcached}{\texttt{kvcached}\xspace}
\newcommand{\eg}{\hbox{\emph{e.g.}}\xspace}
\newcommand{\ie}{\hbox{\emph{i.e.}}\xspace}

%% file: abstract_v1.tex
\begin{abstract}

Inference providers must maintain availability for many LLMs, including low-volume but essential models, making resource efficiency increasingly important as token prices fall. Analysis of production traces reveals a dynamic \emph{bursty-group pattern} in which sets of models become active together and shift over time; existing space- and time-sharing approaches lack principled mechanisms to adapt to this variability, forcing trade-offs between SLO adherence and efficiency. We observe that elastic memory allocation can unify spatial and temporal sharing. Based on this insight, we have developed \tool, a memory-centric LLM co-serving framework that applies \emph{memory ballooning} to reclaim memory across models and support both forms of sharing under a single scheme. \tool's balloon driver, referred to as \codeIn{kvcached}, has been open-sourced at \url{https://github.com/ovg-project/kvcached}, and deployed in production environments across 10K+ GPUs. 

\end{abstract}

%% file: intro_v3.tex
\mysection{Introduction}

Serving LLMs is costly for providers such as Hugging Face and Alibaba Cloud, which must host thousands of base and fine-tuned models, many with low request volumes yet mandatory availability~\cite{aegaeon, slora, xiang2025}. As token prices fall and models vary widely in size and workload patterns, reducing inference cost while preserving performance has become a central objective. The core challenge, however, is not hardware expense but chronic under-utilization: to satisfy strict SLOs, industry practice often dedicates a \emph{model-parallel GPU group} (\ie, one or more GPUs jointly serving a single model instance), guaranteeing immediate responsiveness but wasting substantial capacity. This strategy is particularly inefficient for models with bursty or sparse traffic, and production analyses show that GPU duty cycles commonly fall below 30\%~\cite{aegaeon, xiang2025}.

To improve utilization, prior work has explored various GPU sharing mechanisms at different granularities. 
Because auto-regressive LLM inference is intrinsically memory-bound, tail latency is governed primarily by memory stalls (\eg, page faults, PCIe migrations). 
Moreover, compute sharing is already well studied\textemdash both via hardware mechanisms such as NVIDIA MIG~\cite{nvidiaMIG} and vGPU~\cite{nvidia-vgpu}, and via software solutions~\cite{nvidiaMPS,tally@asplos23,fu2024serverlessllm,lithos@sosp25,gpreempt@atc25,xsched@osdi25}. In contrast, effective memory sharing remains comparatively underexplored. We therefore center this work on advancing effective GPU memory sharing.

The work of memory sharing for LLM serving falls into two categories: space sharing and time sharing. 
Space sharing (\eg, MuxServe~\cite{muxserve}) colocates multiple models on the same GPU to utilize unused memory and compute cycles, which is effective for smaller, low-traffic models that are continuously accessed. Time sharing (\eg, QLM~\cite{qlm} and Aegaeon~\cite{aegaeon}) swaps models in and out of GPU memory, which is ideal for handling sporadic requests to models as the resources can be freed up when models are idle. While these strategies are effective for specific cases, our analysis of a broad range of production serving traces demonstrates that they fail to adapt to more complex and realistic usage scenarios. 

In particular, we have conducted a comprehensive analysis of production traces drawn from major inference providers (\ie, Hyperbolic and Novita AI) as well as independent evaluation platforms (\ie, Chatbot Arena). These traces encompass 11 days to 16 months of traffic across 58 heterogeneous models. Our results indicate that real-world workloads display diverse, rapidly changing patterns that elude any single, fixed sharing mechanism. 

\MyPara{Key observations.} \underline{At the model level}, deployed models show a \emph{bursty-group} behavior: when a burst occurs, requests concentrate on a particular subset of models, and that subset shifts over time, similar to an application’s dynamically changing working set. This pattern is driven by modern compound AI systems and agentic pipelines, which depend on a central reasoning LLM together with a number of fine-tuned or LoRA models for specialized agent tasks. \underline{At the request level}, detailed inspection of the traces confirms that requests are extremely \emph{dynamic and volatile} and fluctuate rapidly and unpredictably, limiting the value of existing GPU-sharing techniques. 
Space sharing, alone, fails to adapt to longer idle periods; for instance, in the Novita AI trace, where models are idle for over 70\% of the time, space sharing creates significant memory fragmentation, leaving 50\% of GPU memory occupied but unused. In contrast, time sharing, alone, struggles with rapid workload fluctuations. The significant latency overhead of swapping (\eg, several seconds) causes severe thrashing during traffic bursts, resulting in a spike of requests violating their latency SLOs due to queuing delays. 

Maximizing resource utilization while preserving SLO attainment requires integrating both strategies: space sharing for colocating low-traffic models and time sharing for burst handling and idle-resource reclamation. However, mingling these approaches is far from trivial due to the fundamentally different challenges they must address\textemdash the flexibility of memory balancing for space sharing vs. the efficiency of model swapping for time sharing. 
One could attempt to switch spatial and temporal sharing strategies dynamically;
yet, co-locating models with conflicting sharing needs creates resource contention. Specifically, a model requiring time-sharing typically relies on preempting resources to maximize burst performance, which directly contradicts the persistent resource availability required by a model colocated through space-sharing. Without a mechanism to fluidly mediate these conflicting operational modes, the system is forced into rigid configurations that cannot adapt to dynamic load changes.

\MyPara{Insight: a memory-centric view to unify space- and time-sharing.} 
Our central observation is that \emph{GPU memory is the unifying bottleneck, and each sharing strategy is concerned with a different portion of that memory}: time-sharing focuses on efficiently swapping in \emph{LLM weights}, whereas space-sharing determines how to scale \emph{KV-cache} capacity among concurrently served models. This perspective parallels \emph{memory ballooning} in virtualization, where a hypervisor dynamically reclaims memory from guest VMs and reallocates it where needed. In the context of LLM co-serving, an analogous “balloon driver” can reclaim memory from idle models to swap in the weights of new models (supporting time-sharing) or from models with low request rates by shrinking their KV-cache reservations (supporting space-sharing). Treating GPU memory as an elastic resource allows the system to fluidly shift between sharing modes and even enable both simultaneously (\ie, apply time-sharing to only a subset of colocated models), ensuring active models obtain the memory required to satisfy their SLOs while maximizing cluster-level efficiency.

\MyPara{\tool.} We developed \tool, a new memory-centric GPU sharing system for multi-LLM serving. At its core, \tool implements the concept of memory ballooning for LLMs to enable \emph{demand-aware, cross-model} memory sharing, a capability missing in existing systems but critical for unifying spatial and temporal strategies. \tool ensures that models can promptly expand their memory footprint to meet latency SLOs, while unused memory is efficiently harvested from others to maximize overall cluster utilization. The design of \tool addresses two key challenges.

\MyPara{Challenge 1: How to enable flexible memory ballooning for LLMs?} Today's LLM serving engines (\eg, SGLang and vLLM) are designed for single-model serving and adopt \emph{per-model static} memory allocation. While techniques like \codeIn{PagedAttention} manage memory efficiently within a model, they operate at the application level and cannot harvest memory across model boundaries.

Flexible memory ballooning requires a mechanism to dynamically resize the physical memory footprint of a model at runtime without disrupting its execution. \tool tackles this through a model balloon driver for serving engines (\S\ref{sec:sys}). 
The model balloon driver decouples virtual and physical GPU memory: serving engines reserve a large virtual address space during initialization, but physical memory is allocated and mapped only on demand. This design enables transparent, fine-grained memory redistribution at a 2\,MB granularity with millisecond-level overhead,
supporting diverse model architectures with zero code change to attention kernels.

\MyPara{Challenge 2: How to place models and schedule requests to maximize SLO attainment?} While the balloon driver provides the mechanism for elasticity, maximizing SLO attainment requires intelligent policies to decide \emph{when} and \emph{where} memory should be allocated. Without proper coordination, contention for shared memory can lead to thrashing and performance degradation.
\tool solves this through decoupled model placement and request scheduling. First, the fast-changing bursty groups (\S\ref{sec:motivation:model}) require a placement algorithm that can quickly decide where each model should run. \tool tackles this with a global demand-complementary placement algorithm that minimizes the KV pressure ratio (KVPR), balancing model memory demand with GPU capacity (\S\ref{sec:sch-global}).
Second, volatile request arrivals and diverse SLO targets require estimating each request’s time slack so the memory manager can steer memory toward latency-sensitive tasks. \tool achieves this via a local slack-aware arbitration algorithm, which leverages a GPU-level request queue to prioritize requests based on their slack (\S\ref{sec:sch-gpu}).

\MyPara{Results.} 
We implemented \tool on top of SGLang~\cite{sglang@arxiv23}, a widely used LLM serving engine. We evaluated \tool comprehensively with two production traces, and with \textbf{58 representative LLMs of varying architectures and sizes, on GPU clusters with up to 32 H100 GPUs. }
Our results show that \tool achieves up to 3.3$\times$ higher time-to-first-
token (TTFT) SLO attainment and $2\times$ higher time-per-output-token (TPOT) SLO attainment given the same number of GPUs.
When targeting the same level of SLO attainment, \tool achieves up to over 2$\times$ cost reduction or 3.5$\times$ more requests compared to the state-of-the-art. 

\tool's balloon driver, referred to as \codeIn{kvcached}, has been open-sourced and actively maintained at \url{https://github.com/ovg-project/kvcached}. \codeIn{kvcached} has been adopted by many industry users to support production workloads in clusters of more than 10K GPUs.
We report usage analyses from two companies at the end of \S\ref{sec:eval}. 

%% file: background.tex
\mysection{Background}
\label{sec:background}

\MyPara{SLOs of Online LLM Serving.}
In online LLM serving, unlike offline batch processing, the system must interact with users or downstream applications in real-time. 
Consequently, providers enforce strict SLOs to guarantee the quality of experience. 
Two key latency metrics define these objectives. 
First, \textit{Time-To-First-Token} (TTFT) measures the latency to process the input prompt and generate the first token, which is a crucial metric for interactive applications, such as chatbots. 
Second, \textit{Time-Per-Output-Token} (TPOT) measures the average inter-token latency during the autoregressive generation, which is essential to ensure that text generation is fluid and matches human reading speeds.

\MyPara{Memory footprint of LLM serving.}
LLM inference is inherently memory-bound, with GPU memory dominated by model weights and KV cache.

\noindent \textbf{\textit{Model weights.}}
LLM weights are massive; for example, a 70B-parameter model requires roughly 140~GB of GPU memory in FP16. 
Under long-tail workloads~\cite{aegaeon}, idle models waste GPU memory; prior work~\cite{aegaeon,fu2024serverlessllm} mitigates this with model eviction and accelerated weight loading.

\noindent \textbf{\textit{KV cache and PagedAttention}.}
The KV cache stores intermediate Key (K) and Value (V) tensors for attention layers to avoid redundant computation. 
Unlike weights, the KV cache is dynamic, growing with sequence length and batch size. 

To improve GPU memory efficiency, mainstream LLM serving systems adopt PagedAttention~\cite{kwon2023efficient}, which splits each request’s KV cache into fixed-size blocks that can be stored non-contiguously and reused across requests.
Frameworks such as SGLang~\cite{sglang@arxiv23} and vLLM~\cite{kwon2023efficient} implement this design by reserving large PyTorch tensors as a KV block pool and mapping each request’s logical blocks to physical offsets in the pool. For each of the $L$ attention layers, a K tensor and a V tensor of shape $(B, T, H, D)$ are allocated, where $B$ is the number of blocks, $T$ is the tokens per block, $H$ is the number of attention heads, and $D$ is the head dimension. The values of $L$, $H$, and $D$ vary across model architectures (\eg, Llama-3-8B ($32, 8, 128$) vs. GPT-OSS-120B ($36, 8, 64$)), resulting in different KV block sizes.

%% file: motivation_v2.tex
\mysection{Production Trace Analysis}
\label{sec:motivation}

\begin{table}[t]\centering
\scriptsize
\begin{tabular}{p{20mm}p{23mm}p{12mm}p{15mm}}\toprule
\textbf{Trace name} &\textbf{Service provider} &\textbf{\# models} &\textbf{Time span} \\\cmidrule{1-4}
Hyperbolic &Hyperbolic~\cite{hyperbolic} &24 &4 months \\\midrule
Novita &Novita AI~\cite{novita-ai} &16 &1 month \\\midrule
Arena-Battle &Chatbot Arena~\cite{chatbotarena@arxiv24} &129 &16 months \\\cmidrule{1-4}
Arena-Chat &Chatbot Arena~\cite{chatbotarena@arxiv24} &84 &11 days \\
\bottomrule
\end{tabular}
\vspace{-2mm}
\caption{Production trace summary.}\label{tab:trace_summary}
\vspace{-1.5mm}
\end{table}

Prior studies have primarily profiled single-model LLM services~\cite{burstgpt, kvcachecache}, but the cross-model dynamics that arise in multi-LLM serving remain largely under-examined. As a result, evaluations in recent multi-LLM serving work often rely on synthetic or simplified workloads, such as Poisson/Gamma arrivals~\cite{muxserve, qlm, aegaeon}, temporally rescaled traces from non-LLM domains~\cite{weaver-atc25, fu2024serverlessllm}, or dataset-driven replays~\cite{sharegpt}. 

To address these gaps, we conduct a detailed characterization of production workloads from major inference providers (\ie, Hyperbolic and Novita AI) as well as independent evaluation platforms (\ie, Chatbot Arena), explicitly capturing the fine-grained, time-varying dynamics of multi-LLM traffic. The four traces span 24–129 deployed models over
weeks to months of traffic, summarized in Table~\ref{tab:trace_summary}. 
In this section, we utilize the traces from Novita AI as a representati4ve example to illustrate our observations and report cross-trace statistics. 
We include a detailed CDF analysis in Appendix~\ref{ap:trace}.

\textbf{\begin{figure}[t]
    \centering
    \includegraphics[width=0.8\linewidth]{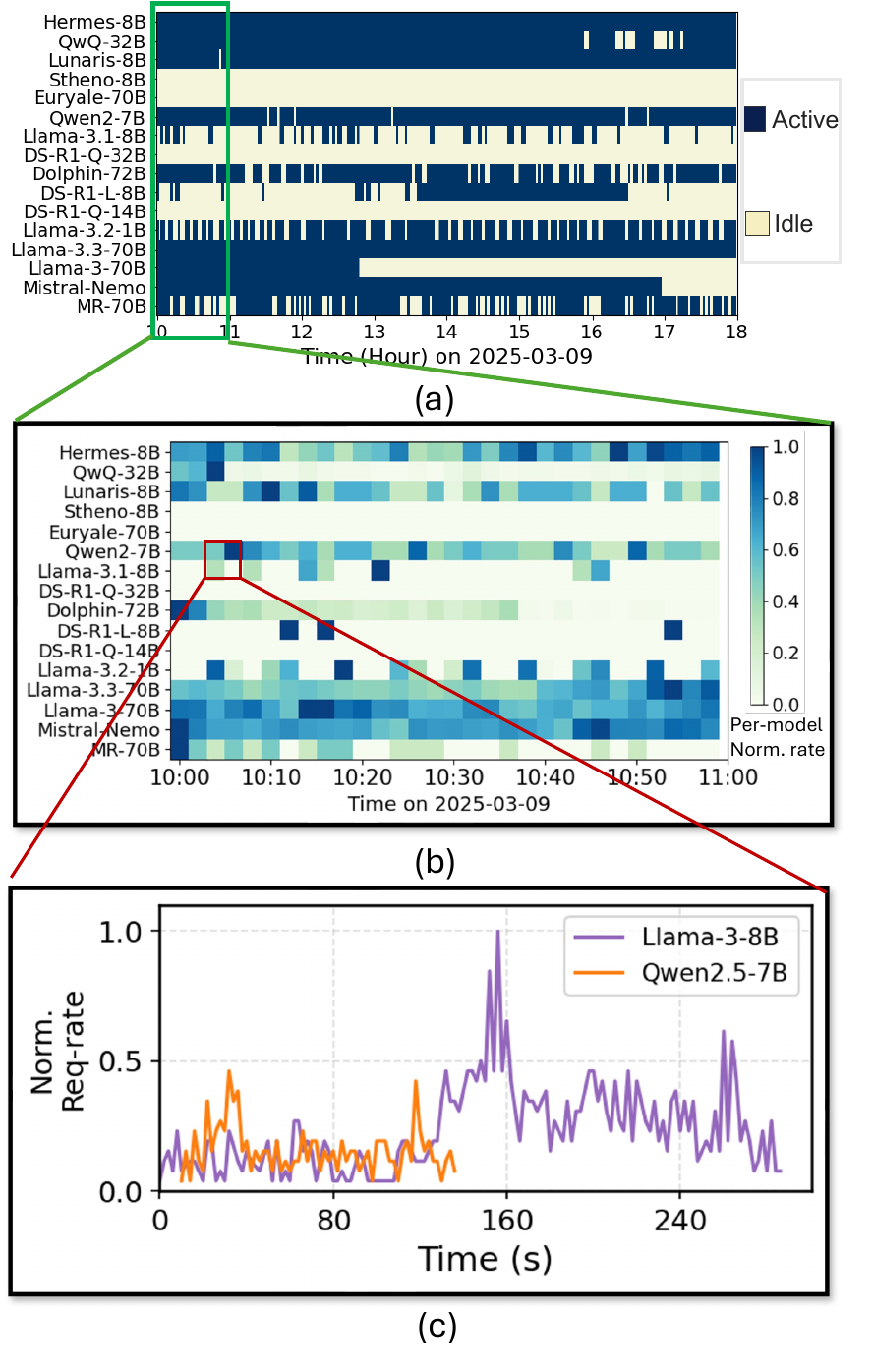}
   \vspace{-4mm}
    \caption{Combined view of model-request dynamics; the x-axes denote time. (a) Variation of concurrently active model groups over a six-hour window (8am–4pm): the y-axis lists all models from the Novita trace; each cell represents a three-minute interval, with dark/light shading indicating whether the model is active. (b) Zoomed-in view of volatile request patterns from 11am–1pm, showing per-model normalized request rates, where darker colors indicate higher rates.  (c) A further zoom-in over a 5-minute window from (b), comparing two models on a shared normalization.}
    \label{fig:working-set}
  \vspace{-1.5mm}
\end{figure}}

\mysubsection{Model Level: Shifting Bursty Groups}
\label{sec:motivation:model}
We begin by examining the dynamics of concurrently active models (\ie, those processing at least one request) from the Novita trace over a six-hour span in Figure~\ref{fig:working-set}(a), where a horizontal slice of a model represents its active/idle activities over time, and a vertical slice at a point of time reveals which subset of the models is active at the point. 
Our analysis reveals two properties that are common for production workloads:  \emph{bursty groups} and \emph{heterogeneous access patterns}.

\noindent \textbf{Vertical behavior: bursty groups.}
Vertically, deployed models demonstrate a \textit{bursty-group} behavior: models receive requests in short, irregular bursts separated by long and variable idle intervals. Because these bursts occur asynchronously across models, only a small subset is active at any moment, and this subset changes rapidly over time. We refer to this transient set as the \textit{bursty group}, analogous to how an application's \emph{working set} changes during its execution\textemdash there are common models that appear in almost every group but most models experience sporadic requests and hence appear only in a small number of groups. 
Across the four traces, 23\%--50\% of models are active concurrently on average, and the active set changes 54--766 times per hour.
Although the total number of deployed LLMs can be large, only the bursty group drives actual demand. Efficient GPU utilization thus requires adapting resources to this shifting model group, rather than allocating resources statically to all models.

\noindent \textbf{Horizontal behavior: Heterogeneous activation patterns.}
Horizontally, the workload exhibits significant heterogeneity across models in activation patterns.
Some models, such as \code{Llama-3.3-70B}, sustain long stretches of continuous activity, while others (like \code{Distilled-Deepseek-R1-Qwen-14B(DS-R1-Q-14B)}) activate only in sparse, short-lived instances with only a small group of applications/users using it.
This heterogeneity is largely driven by the diverse roles that models play in modern compound AI systems and agentic pipelines. Large general-purpose models (\eg, \code{Llama-3.3-70B}) often act as central reasoning or planning components and remain active throughout user interactions, leading to sustained request streams. In contrast, smaller distilled models (\eg, \code{DS-R1-Q-14B}) are typically invoked as auxiliary components—such as for tool use or verification—and are triggered only at specific pipeline stages, resulting in intermittent, bursty activations.

Aggregating requests across many applications and users yields a mixed access pattern in which some models remain persistently active, others appear only sporadically, and even the same model may exhibit different behaviors over time. These differences have direct implications for resource management: long-running models benefit from sustained colocation, whereas short and scattered bursts favor opportunistic execution. Consequently, multi-LLM serving systems must adapt to the heterogeneous activation behavior of each model, rather than assuming uniform access patterns.

\noindent \textbf{Implication: Inefficiency of static partitioning.} The existence of shifting \textit{bursty groups} fundamentally challenges any static strategies. Since only a small, changing subset of models drives demand at any given moment, statically partitioning GPU memory across all models (to ensure availability) results in severe resource fragmentation. In the Novita AI trace, for example, models are idle for over 70\% of the time on average. Under a static memory-partitioning regime, the memory reserved for these idle models is wasted, preventing active models in the current bursty group from utilizing that capacity to scale up and meet their SLOs.

\mysubsection{Request Level: Dynamic and Volatile}
\label{sec:motivation:request}

We next zoom in to a shorter period of time to examine the fine-grained dynamics of individual models.
Figure~\ref{fig:working-set}(b) depicts the request rates for the models over a two-hour window, normalized within each model so that low-volume bursts remain visible alongside high-volume ones.

\noindent \textbf{Extreme volatility and pattern shifts.}
As shown in Figure~\ref{fig:working-set}(b), request streams shift patterns frequently.
Even models that appear continuously active in coarser views actually exhibit rapid oscillations between low-demand periods and short, intense bursts (\eg, \code{Qwen2-7B}).
Similarly, models that are only sporadically invoked display burst sizes and inter-arrival times that vary significantly over time (\eg, \code{Llama-3.1-8B}).

\noindent \textbf{Abrupt pattern transitions.} 
Furthermore, these workload patterns often undergo rapid, unforeseen transitions. Figure~\ref{fig:working-set}(c) provides a granular, 5-minute zoom-in view of request arrivals for two representative models. This window reveals a stark shift in interaction dynamics: a phase of moderate, interleaved activity\textemdash where both models contend for resources simultaneously\textemdash abruptly gives way to a concentrated burst from a single model. Such sudden pattern shifts occur without discernible triggers, implying that the system cannot rely on historical stability or long-term pattern recognition; instead, it must possess the agility to react to instantaneous changes. 
Such transitions are pervasive: across our traces, many models exhibit a request rate coefficient of variation above 1 and 40–100 idle intervals per hour.

\noindent \textbf{Implication: Instability of pure time sharing.} The extreme \textit{volatility} and rapid pattern transitions undermine pure time-sharing strategies. As seen in Figure~\ref{fig:working-set}(c), request streams frequently transition into phases of interleaved activity where multiple models are active simultaneously. In these scenarios, a time-sharing system relying on swapping must constantly evict and reload weights to serve alternating requests. This leads to model thrashing, where the latency overhead of loading weights dominates execution time, causing immediate SLO violations. This volatility necessitates the stability of space sharing, where working sets remain resident to handle interleaved demand without swapping penalties.

\mysubsection{Applying Existing Approaches}
\label{sec:motivation:sota}

\begin{figure}[t]
  \centering
  \begin{subfigure}[t]{0.49\linewidth}
    \includegraphics[width=\linewidth]{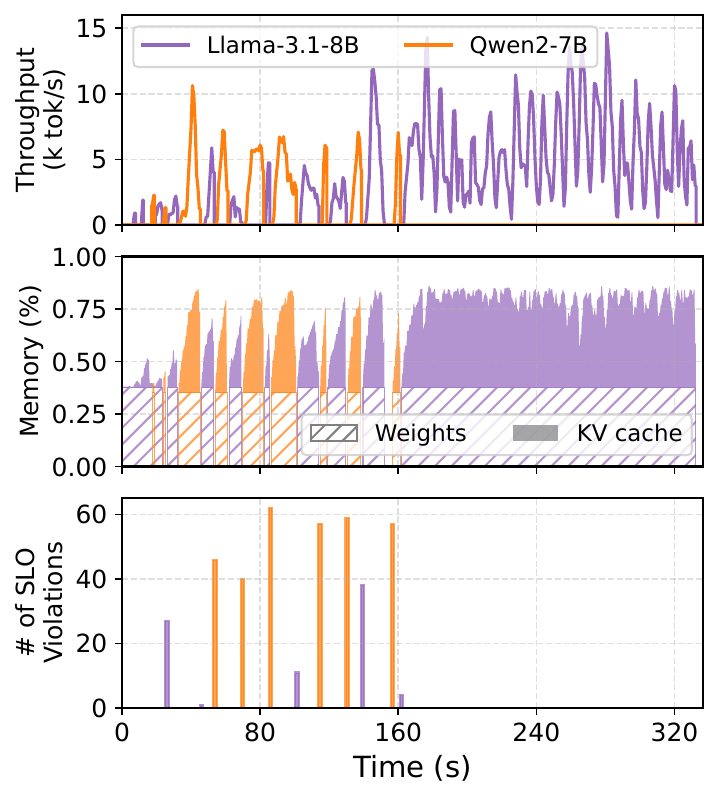}
    \caption{Pure time sharing.}
    \label{fig:motivation-time}
  \end{subfigure}
  \begin{subfigure}[t]{0.49\linewidth}
    \includegraphics[width=\linewidth]{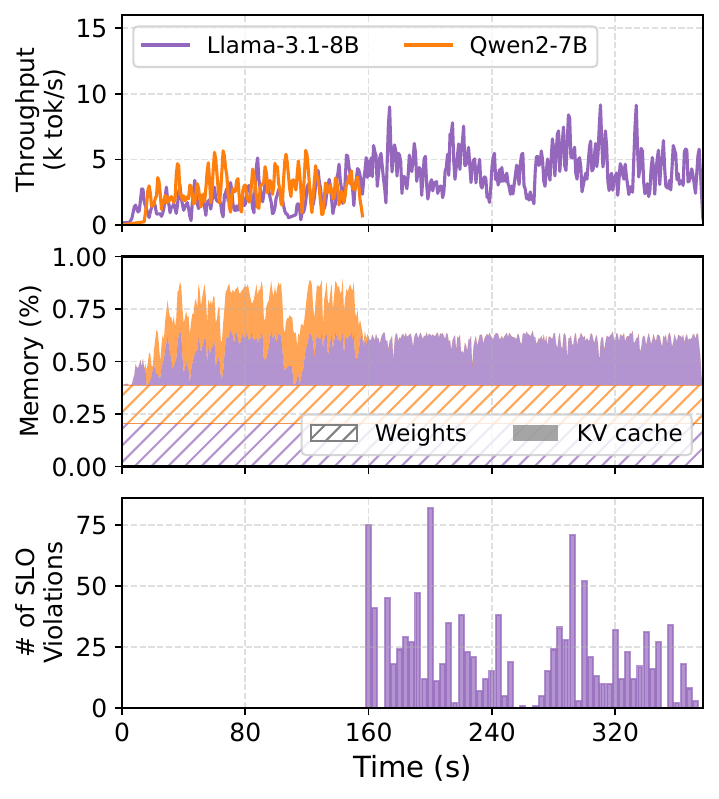}
    \caption{Pure space sharing.}
    \label{fig:motivation-space}
  \end{subfigure}
  \caption{Performance of pure time- (a) and space-sharing (b) for the trace in Figure~\ref{fig:working-set}(c). Pure time sharing in (a) causes model switching when both models have requests, while pure space sharing in (b) causes queuing delay during requests spike.}
  \label{fig:motivation-exp}
\end{figure}

To understand the implications of these abrupt pattern transitions, we ran two representative state-of-the-art strategies, \textit{time sharing} (QLM~\cite{qlm}) and \textit{space sharing} (static partition~\cite{nvidiaMIG, fractionalGPUs}), on the trace segment shown in Figure~\ref{fig:working-set}(c).
Figure~\ref{fig:motivation-exp} illustrates the resulting memory usage and cumulative SLO violations for each approach.

\MyPara{Failure of time sharing.}
Time sharing manages resources by swapping models in and out of GPU memory.
This approach struggles during the initial phase of \textit{interleaved activity}.
As both models receive frequent, overlapping requests, the system is forced to repeatedly evict and reload weights to serve alternating inputs.
As shown in Figure~\ref{fig:motivation-time}, this leads to severe \textit{model thrashing}: the GPU spends more time on PCIe transfers and engine re-initialization than on computation.
As such, request queues grow rapidly, resulting in a spike of SLO violations even before the workload intensity peaks.

\MyPara{Failure of space sharing.}
Space sharing avoids swapping by partitioning GPU memory between models.
While this provides stability during the interleaved phase, it fails catastrophically when the workload transitions to the burst phase, as shown in Figure~\ref{fig:motivation-space}.
Although one model becomes idle, its model weights remain locked, preventing the active model which is now experiencing a demand surge from utilizing the full GPU capacity.
This rigid isolation creates artificial resource scarcity: the active model is starved of KV cache memory while adjacent memory sits idle.
The result is a throughput cap that causes significant queuing delays and SLO violations during the burst.

\mysubsection{Takeaway: A Hybrid Approach} \label{sec:motivation:takeaway}

Our analysis demonstrates that production multi-LLM workloads exhibit conflicting requirements that no single existing strategy can satisfy: 
\begin{itemize}[leftmargin=1em,nosep]
\item \textbf{Bursty groups benefit from time-sharing capabilities:} At the model level, the active set shifts over time. Pure space sharing fails here because it rigidly locks memory for idle models, causing fragmentation and preventing active models from scaling. The system must be able to evict idle resources to maximize utilization. 
\item \textbf{Volatile requests benefit from space-sharing capabilities:} At the request level, demand fluctuates rapidly and often interleaves across models. Pure time sharing fails here because swapping causes severe thrashing during these volatile periods. The system must be able to keep active models colocated in memory to provide low-latency access. \end{itemize}

\noindent \textbf{Motivation for \tool.}  These two patterns capture the workload at coarse- and fine-grained levels, respectively. Because they reflect different facets of the same underlying behavior, they cannot be addressed in isolation, necessitating a hybrid approach: the system must fluidly shift between the strategies at runtime. It needs the \textit{elasticity} to act like a time-sharing system when reclaiming memory from the idle models, while simultaneously acting like a space-sharing system to co-serve the models in an active bursty group without causing SLO violations. This calls for a design that manages GPU memory as a dynamic, shared resource rather than static partitions.

%% file: overview.tex
\mysection{Overview}

\begin{figure}
    \centering
    \includegraphics[width=0.9\linewidth]{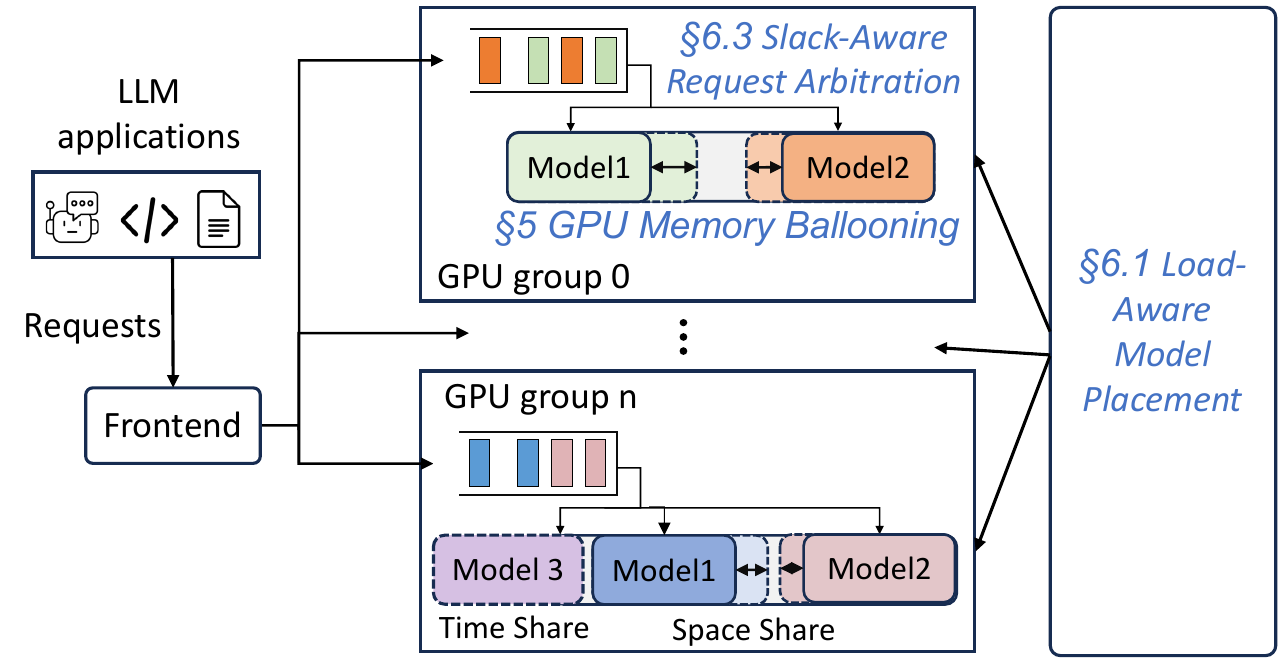}
    \caption{The system architecture and design overview of \tool.}
    \vspace{-0mm}
    \label{fig:design-overview}
\end{figure}

\tool is a memory-efficient GPU sharing system for cost-effective multi-LLM serving.
Figure~\ref{fig:design-overview} shows its system architecture and design overview. \tool receives inference requests at its frontend, which routes them to the corresponding LLMs for processing. 
To improve cost efficiency, \tool serves LLMs with flexible combinations of space and time sharing. For instance, high-rate models may occupy GPUs exclusively, while multiple low-rate models can be colocated on a single GPU or a model-parallel GPU group. In \tool, a GPU group represents the strict scheduling boundary\textemdash a set of GPUs tightly coupled to jointly serve one model instance. \tool's global scheduler treats each of these groups as a distinct, indivisible resource unit. Idle models are temporarily evicted to CPU DRAM and reactivated when new requests arrive. \tool continuously adjusts its sharing strategies based on runtime workload to maximize overall SLO attainment. 

\tool achieves this with two key designs for flexible, demand-aware cross-model memory sharing and coordination. First, it introduces a balloon driver, referred to as \codeIn{kvcached} (\S\ref{sec:sys}) that enables flexible and efficient memory sharing between models, forming the foundation for rapid adaptation to workload variations and policy changes.
Second, \tool optimizes the overall resource allocation through a memory-centric control plane that uses a combination of (1) load-aware model placement (\S\ref{sec:sch-global}) that maximizes memory headroom for GPUs to facilitate ballooning and (2) slack-aware request arbitration that prioritizes requests on each GPU for maximizing SLO attainment (\S\ref{sec:sch-gpu}).

\MyPara{Relation with autoscaling.}
\tool is orthogonal to and can be seamlessly integrated with autoscaling~\cite{blazescale, sgl‐project-ome, ai-dynamo_2025, llm-d_2025}.
\tool focuses on efficiently sharing GPUs between models (replicas) while autoscaling automatically scales up/down the number of model replicas.

%% file: design-v2.tex
\mysection{GPU Memory Ballooning for Multi-LLMs}
\label{sec:sys}

This section describes how \tool enables flexible cross-model memory sharing with our balloon driver, providing a foundation for adapting policies to dynamic workloads.

\mysubsection{Requirements and Challenges}
\label{sec:flex-mem}

Mainstream LLM serving engines~\cite{sglang@arxiv23,kwon2023efficient} adopt block-based KV token management (\eg, PagedAttention~\cite{kwon2023efficient}). 
While this design effectively reduces \emph{request-level} memory fragmentation \emph{within a single} LLM, it requires the application (serving engine) to manage the GPU physical memory directly, by pre-allocating a large portion of GPU memory as the KV block pool.
This memory thus remains occupied even when there is no active request, and the KV blocks within the tensor are unused.
This prevents the redistribution of unused memory across different LLMs, limiting the ability to support flexible time- and space-sharing among multiple LLMs. 
This leads to four key requirements for \tool.

\MyPara{R1: Fast memory reallocation between model weights and KV cache.}
When a model is swapped out, the memory released by its weights should be quickly repurposed by other models as KV cache. Similarly, when an idle model is reactivated upon new requests, the running models must promptly release unused KV cache memory for the new model's weights and KV cache.
However, this is hard with statically reserved KV pool tensors, whose physical GPU memory cannot be partially released.
A workaround is to copy active token blocks into a smaller tensor and then destroy the old one. 
However, this not only requires waiting for the slow copy to complete, but also demands enough GPU memory to hold both tensors during the transient period.

\MyPara{R2: KV cache sharing across diverse model architectures.}
For models colocated on a GPU, their KV cache memory should be flexibly shared; \eg, unused KV cache memory from one LLM can be easily reused by others.
However, LLMs often have diverse KV cache layouts, resulting in different tensor shapes, \eg, different head dimensions and number of attention heads. 
Therefore, simply using a pool of unified KV tensors across models is infeasible: different token sizes lead to misaligned tensor shapes, and different layer counts cause inconsistent numbers of tensors.

\MyPara{R3: Minimal redistribution overhead and memory fragmentation.} 
Since memory redistribution may occur frequently, it must incur low overhead while minimizing fragmentation to improve utilization. 
This rules out a strawman design that splits the reserved KV pool tensor into multiple smaller segments so that free segments from one model can be reused by another. 
This is because choosing the segment size is inherently challenging: too many small segments increase second-level indexing overhead, while too few large segments cause severe fragmentation.

\MyPara{R4: Transparent to existing serving frameworks.} 
The new mechanism should remain fully compatible with existing serving frameworks, where the KV block pool is allocated as large tensors, and require no changes to attention kernels. 
The previously mentioned segment-based method violates this requirement, as it splits the tensor into segments, which necessitates extensive kernel modifications. 

\MyPara{Summary.} 
Multi-LLM serving must efficiently cope with heterogeneous memory demands (\eg, different model weights and KV token blocks), each with distinct semantics (\eg, head dimensions and layer counts) and lifecycles. 
Existing single-LLM systems manage GPU memory entirely at the application level, imposing their own semantics on top of a fixed GPU allocation. 
This approach works well for a single model, which can efficiently manage its memory in isolation. 
In multi-LLM settings, however, such isolation becomes a barrier, as memory needs to be shared and reallocated across models dynamically to accommodate bursty group behavior and workload volatility.

\begin{figure}
    \centering
    \includegraphics[width=0.99\linewidth]{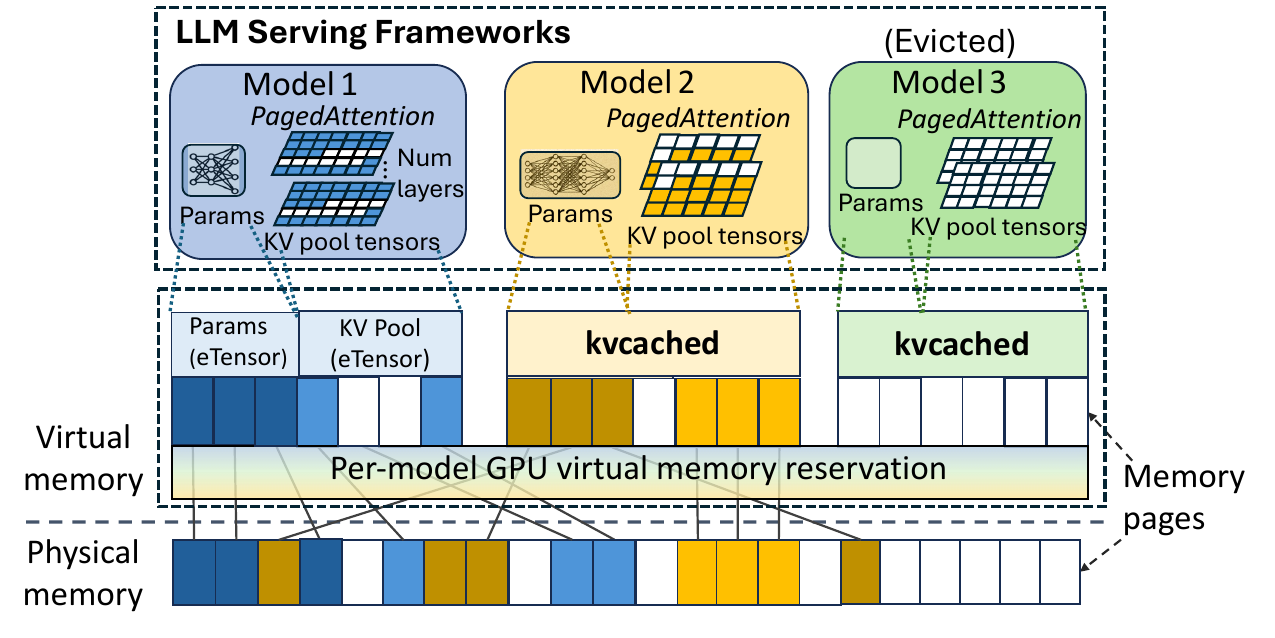}
    \caption{Memory ballooning for effective multi-LLM serving.}
    \vspace{-0mm}
    \label{fig:elastic-engine}
\end{figure}

\mysubsection{Solution: GPU Memory Ballooning}

To meet all the above requirements,
our key insight is that memory management should be \emph{pushed down to the GPU runtime level.}
At this level, the system can treat all models uniformly, allowing GPU memory to be transparently redistributed across models, while applications manage only the memory allocated to them with their own semantics.

Building on this idea, \tool introduces a \emph{balloon driver}, named \codeIn{kvcached}, which sits as a shim between inference engines and GPU memory. 
Open sourced at \url{https://github.com/ovg-project/kvcached}, \codeIn{kvcached} manages the entire GPU physical memory for LLM serving, including both model weights and the KV cache pool.
Meanwhile, \codeIn{kvcached} reserves a large contiguous virtual address space for each engine, which presents it to the engine as if it were dedicated GPU memory. 
On top of this reserved space, \codeIn{kvcached} translates application-level semantics (\eg, weights, KV caches, and intermediate buffers) into allocations of physical GPU memory pages. 
Physical pages are created only on demand and mapped to virtual addresses lazily,
allowing memory to expand and shrink as workloads change, as illustrated in Figure~\ref{fig:elastic-engine}. 

Specifically, \codeIn{kvcached} meets all the requirements in \S\ref{sec:flex-mem} through the following designs.

\MyPara{D1: Unified model weights and KV cache management.} 
\codeIn{kvcached} achieves fast memory redistribution between model weights and KV cache through unified virtual and physical memory management. 
Since \codeIn{kvcached}-managed memory is agnostic to application semantics, both weights and KV cache can be seamlessly reallocated across engines.
Moreover, \codeIn{kvcached} dynamically adjusts physical memory 
limits: when an evicted model is reactivated, it shrinks the limits of other models on the same GPU, bounding their allocations and immediately freeing space for the new model.
The opposite happens when a model is evicted.

\MyPara{D2: Automatic token block mapping.} 
To support models with different layer counts and token sizes, \codeIn{kvcached} employs an internal KV cache manager to map token blocks onto underlying virtual and physical GPU pages. \tool assigns token blocks to available slots within a virtual page, or allocates a new physical page when necessary. To prevent size conflicts and ensure portability across architectures, the KV cache manager further segregates token blocks from different models onto distinct memory pages.

\MyPara{D3: Overhead and fragmentation optimizations.} 
\codeIn{kvcached} minimizes memory redistribution overhead and fragmentation through three key optimizations.
First, mainstream serving frameworks maintain separate K and V tensors per layer, 
requiring $2L$ page allocations across all layers each time. To reduce this overhead, \codeIn{kvcached} reorganizes the memory layout in virtual space, so that all layers' K and V vectors of a token are stored in contiguous virtual space, requiring only one batch allocation for all pages ($2L\times$ speedup). 
Second, \codeIn{kvcached} uses a pre-allocation thread to asynchronously prepare a small buffer of GPU pages. Engines draw new pages from this buffer first, while released pages are returned to it and only physically freed if the buffer exceeds its limit or memory must be reclaimed for a new model.
Finally, to reduce fragmentation, \codeIn{kvcached} uses 2MB memory pages, and prioritizes using partially filled pages.

\MyPara{D4: Elastic tensor abstraction.} 
\codeIn{kvcached} ensures transparency with existing LLM serving frameworks by introducing an \textit{elastic tensor (eTensor)} abstraction, implemented via PyTorch's extension interface
to abstract away the details of physical page allocation from the serving engine. 
Elastic tensors can be used exactly like regular PyTorch tensors, requiring no modifications to attention kernels and remaining fully compatible with CUDA graph optimizations commonly used in LLM serving frameworks.

\mysubsection{Fast Model Loading}
\label{sec:model-loading}
Model swapping speed directly impacts the flexibility of GPU sharing. High latency can hinder the timely swapping of models with strict SLOs, limiting policy adaptability.
While deactivation is straightforward, \ie, terminating the engine and releasing all memory, reactivation is more complicated, which involves:
(1) initializing a new serving engine and reserving a new virtual address space for KV cache pool; and
(2) loading the model weights from CPU DRAM.
If done \naively, this process can take tens of seconds---far exceeding the TTFT SLOs of online LLM inference, which are often within a few seconds or less.

\MyPara{Reusable engine pools.}
The root cause of (1) lies in the tight coupling between the engine and the model it serves. In current systems, an engine shares the same lifecycle as its model---when a model is evicted, its engine is also terminated, along with its virtual address space. As a result, every model activation incurs the full cost of engine initialization.
\\
\tool eliminates this overhead by decoupling the lifecycles of the engines and models. Specifically, it maintains an engine pool on each GPU, where engines are pre-initialized with virtual address space and distributed contexts. Upon model activation, \tool selects an available engine from the pool and starts model loading directly. When a model is evicted, its physical memory is released, but its engine with virtual address space is returned to the engine pool for future reuse.
\\
However, an engine cannot directly reuse previously reserved virtual memory space to serve a new model. 
This is because current inference engines perform index-based token access, which depends on a model-specific memory layout that is incompatible with models of different architectures, \eg, different numbers of layers or token sizes.
To address this, \tool introduces a KV cache virtual memory manager to manage the pre-reserved virtual memory spaces in the engine pool. When a new model is loaded, the manager dynamically aligns the reserved virtual space to match the memory layout required by the new model (one-time effort), and then creates a new \kvcached based on the aligned memory spaces.  
The \kvcached can then correctly and efficiently locate the virtual memory page where each token resides during inference. 
\\
\MyPara{Parallel model weight loading.}
The time spent on (2) model weight loading is heavily influenced by the utilization of the CPU–GPU interconnect bandwidth. 
We found that loading models \naively via the \texttt{cudaMemcpyAsync} API to a GPU fails to saturate the interconnect bandwidth, even when invoked from multiple threads. This may be due to all \texttt{cudaMemcpyAsync} operations targeting the same GPU executing serially, limited by the CUDA driver and hardware.
\\
\tool overcomes this bottleneck by chunking model weights into smaller segments, loading them in parallel across multiple GPUs on the same node, and then aggregating them to the target GPU via high-speed NVLink interconnects. This parallelized strategy significantly accelerates model loading.
To minimize interference with running workloads on GPUs, \tool partitions model weights at the granularity of individual weight tensors and loads them in a streaming fashion. As a result, each GPU only needs to maintain a small buffer (\eg, 30MB), minimizing possible memory contention.

%% file: scheduler-v3.tex
\mysection{A Memory-Centric Control Plane}\label{sec:scheduling}

The memory ballooning mechanism in \S\ref{sec:sys} provides the elasticity needed to unify spatial and temporal sharing, but leveraging it effectively requires policies that adapt to evolving bursty groups and volatile request patterns. Our goal is to improve memory efficiency and maximize SLO attainment for TTFT and TPOT through coordinated memory sharing. Because TTFT and TPOT can interfere~\cite{zhong2024distserve}, we prioritize TTFT, whose prompt length is known, while noting that the resulting improvements in resource scheduling also benefit TPOT by reducing memory-induced preemptions.

\MyPara{Challenges.}
Unifying time and space sharing complicates scheduling because decisions must jointly span models, requests, and tokens, all of which shape GPU memory usage and latency. These layers are tightly coupled: model residency sets memory availability, request concurrency determines consumption, and token execution affects both utilization and SLOs. Optimizing any one dimension in isolation—e.g., maximizing colocation or strictly protecting SLOs—can destabilize the others. Consequently, time- and space-sharing choices form a large, interdependent decision space where placement, memory pressure, and request scheduling continually interact in non-trivial ways.

A natural formulation of this problem is a joint optimization over model placement, request scheduling, and memory allocation (\eg, via an ILP), but this quickly becomes intractable. The placement space alone is enormous: with 
$M$ models and $N$ GPUs, there are N$^M$
 possible assignments even before considering migration decisions. Request scheduling is likewise combinatorial, as differing prompt lengths, arrival times, SLOs, output lengths, and engine states yield a factorial explosion in possible execution orders. Moreover, placement and scheduling are tightly coupled—migrating even one model changes which GPUs can serve which requests and how KV caches evolve, thereby altering future scheduling decisions. All of this must be decided online, with the scheduler observing only recent arrivals while lacking knowledge of future bursts, output lengths, or concurrency.

\MyPara{Insight.} Our key insight is that \emph{this complexity simplifies when managed through a memory-centric lens}. Models, requests, and tokens influence system behavior primarily through how they consume and contend for GPU memory, while SLO attainment is directly governed by the availability of that memory at execution time. By making memory the central optimization target, we unify multiple interacting objectives, utilization efficiency, latency control, and stability, under a single principle: controlling memory contention. This reduces a multi-metric, multi-level scheduling problem into a tractable design centered on memory management.

Memory contention manifests at multiple scopes: cluster-level model activity drives coarse-grained memory demand, while fine-grained request dynamics determine SLO outcomes. No single control loop can manage both effectively, motivating a memory-centric hierarchical scheduler: a macro, cluster-level plane that shapes memory pressure via model placement, and a micro, GPU-level plane that allocates memory per request to satisfy heterogeneous SLOs. Coordinating these layers enables \tool to maintain high utilization and robust SLO attainment.

\mysubsection{Load-Aware Model Placement} \label{sec:sch-global}

The objective of the global model placement strategy is to place models across GPUs to maximize the headroom available for memory ballooning.  If too many models from the same active bursty group are colocated, they will simultaneously demand memory for KV-cache expansion, leading to severe contention and out-of-memory errors.

To prevent this, \tool employs a load-aware placement strategy. Because predicting exact memory usage is difficult due to unknown output lengths of LLM inference requests, we utilize a robust heuristic referred to as the KV Pressure Ratio (KVPR). KVPR quantifies the intensity of memory contention on a GPU by comparing the aggregate urgency of memory demand against the remaining capacity, calculated as $\frac{w\_token\_rate}{shared\_kv}$, where $w\_token\_rate = \frac{token\_rate * token\_size}{SLO}$ represents the SLO-weighted token memory usage rate of a model,
indicating its memory demand per unit time. By counting both input tokens from newly admitted requests and decode tokens produced by running requests per unit time, $token\_rate$ captures the full KV-cache growth rate and helps KVPR accurately reflect GPU memory pressure. We use the TPOT SLO for $SLO$ because autoregressive decoding dominates generation latency and is especially sensitive to KV-cache memory contention. 
$shared\_kv$ is the memory available for the KV cache on a GPU.  A high KVPR indicates a high-pressure GPU where memory ballooning is likely to be stifled. 

Algorithm~\ref{alg:global} realizes our model placement strategy. \tool prioritizes the most aggressive memory consumers by sorting models in descending order of their SLO-weighted token usage rates $w\_token\_rate$, ensuring that high-demand workloads are allocated resources first (Lines 1–3). For each model, the scheduler identifies the GPU that minimizes the resulting KVPR. This step inherently enforces complementarity: by targeting the device with the lowest existing pressure, \tool effectively colocates high-demand models with low-demand ones, thereby maximizing the capacity available for the high-demand model to balloon its memory usage (Lines 5–8). 
If the selected GPU differs from the model's current GPU, \tool migrates the model accordingly. However, this migration incurs overhead from engine switching and model weight loading.
To avoid unnecessary migrations with marginal benefit, \tool compares the KVPR of the best and current GPUs, and proceeds only if the improvement exceeds a threshold $\tau$ (Line 8).
Finally, \tool assigns the model to its selected GPU and updates the corresponding GPU states (Lines 9–12).

\MyPara{Analysis.} This greedy approach approximates the optimal solution by bounding the maximum KVPR across the cluster, ensuring that no single GPU becomes a disproportionate bottleneck that stifles memory sharing. A detailed formal analysis is provided in Appendix~\ref{ap:alg-gloabl}. The algorithm also seamlessly supports large models utilizing Tensor Parallelism (TP). By treating each TP partition as a distinct scheduling unit with anti-affinity constraints, \tool ensures they are placed on separate GPUs, thereby aggregating memory capacity across devices to satisfy massive KV demands. 

\MyPara{Model eviction and activation.}
The placement strategy evicts a model if it remains idle and GPU resources become constrained for other models. \tool performs eviction based on an empirical threshold, which can be determined by analyzing the idle interval distribution; a detailed sensitivity analysis is provided in Appendix~\ref{sec:sensitivity}.
When the model receives new requests, \tool immediately reactivates it by placing it on the GPU with the lowest KV pressure ratio, drawing from a pre-warmed engine pool with reusable distributed contexts and buffers to avoid re-initialization, and using parallel weight loading to keep cold-start latency within TTFT budgets (\S\ref{sec:model-loading}).

\MyPara{Model migration.}
Model migration is designed to preserve TTFT during placement changes. Rather than stopping request processing while the destination instance is prepared, \tool keeps the source instance active and continues serving requests until the target is ready. This allows migration latency to be overlapped with ongoing execution, so requests experience only the short switch-over delay. 
Therefore, \tool's effectiveness does not depend on fast interconnects. Practically, when the source and target GPUs are connected via NVLink, \tool uses NVLink as an optional shortcut for migrating model weights and resident KV caches. In non-NVLink environments, \tool leverages GPUDirect RDMA~\cite{gpudirect-rdma} if available, or falls back to standard eviction and reactivation, which takes at most a few seconds for <70B models.

\begin{algorithm}[t]
\small
\caption{Model Placement to Maximize Memory Headroom.}
\label{alg:global}
\begin{algorithmic}[1]
\Require 
    Number of GPUs $N$, GPU memory capacity $C$, migration threshold $\tau$, and $M$ models. Each model $m_j$ has: token rate $t_j$, weight $w_j$, token size $tz_j$, current device index $g_j$, and latency SLO $s_j$.

\Ensure Assign each model to a GPU to balance the resource demand and remaining memory capacity.

\State Sort models by \red{$\frac{t_j * tz_j}{s_j}$} in descending order. Denote the sorted sequence as $m_1, m_2, \dots, m_M$. 

\For{$i=1$ to $N$}
  \State $shared\_kv_i \leftarrow C$; $w\_token\_rate_i \leftarrow 0$
\EndFor

\For{$k=1$ to $M$}
  \State \textcolor{blue}{/* find the GPU $best\_idx$ that minimizes KVPR */}
  \State   $best\_r,\ best\_idx \leftarrow \text{(min, argmin)}\frac{w\_token\_rate_i}{shared\_kv_i}$
  \State   $current\_r \leftarrow \frac{w\_token\_rate_{g_k}}{shared\_kv_{g_k}}$
  \State   $best\_gpu \gets 
  \begin{cases}
    best\_idx, & \text{if } current\_r - best\_r > \tau \\
    g_k,       & \text{otherwise}
  \end{cases}$

  \State Assign model $m_k$ to $best\_gpu$
  \State $w\_token\_rate_{best\_gpu} \gets w\_token\_rate_{best\_gpu} + \frac{r_k}{s_k}$
  \State $shared\_kv_{best\_gpu} \gets shared\_kv_{best\_gpu} - w_k$
\EndFor
\State \Return Model-to-GPU placement
\end{algorithmic}
\end{algorithm}

\subsection{Slack-Aware Request Arbitration} \label{sec:sch-gpu}

While global placement balances demand, memory contention arises locally when resident models process requests concurrently. Under unified sharing, strict isolation causes fragmentation, whereas uncontrolled sharing risks starvation, with urgent requests blocked by long-running ones. Models on the same GPU may compete for KV cache memory, and without coordination a high-rate, relaxed-SLO model can monopolize memory and degrade SLO attainment for stricter workloads.

A \naive approach is to cap each model’s memory usage, but choosing appropriate limits is difficult under dynamic workloads and heterogeneous SLOs: conservative caps throttle throughput, while generous ones starve other models. Adapting limits at runtime helps but cannot take effect immediately, since memory must be freed by completing in-flight requests—a process that can take seconds depending on request length and load. This coordination challenge arises because each engine maintains its own queue and greedily schedules requests as memory becomes available.

To resolve this, \tool employs a slack-aware request arbitration strategy. Instead of maintaining separate queues per model, \tool uses a shared per-GPU request queue that arbitrates access to each GPU's physical memory pool to coordinate the memory usage for heterogeneous SLO requests. Our goal is to prioritize requests that are most critical for maximizing SLO attainment. \tool goes beyond simple deadline prioritization by leveraging the exact time slack of each request, which is defined as the buffer between a request's deadline and its required execution time.

With both the deadline ($d_r$) and execution cost ($e_r$) known, \tool transforms the scheduling challenge into a classic deadline problem. We adopt the Moore-Hodgson algorithm~\cite{moore1968n}, which minimizes the number of deadline misses.
As shown in Algorithm~\ref{alg:local}, given a set of requests $R$, \tool first sorts them in ascending order of their prefill completion deadlines (Line 1).
Then, for each request in the sorted list, \tool appends it to the schedule list $S$ and checks whether it can finish within its TTFT SLO (Lines 3-7). 
If the most recently added request cannot meet its deadline, \tool evicts the request in $S$ with the longest execution time (Lines 9–11). It then moves to the next request and continues this process until evaluating all requests.
Finally, \tool dispatches the accepted requests in $S$ following their order in the schedule.

\begin{algorithm}[t]
\small
\caption{GPU-Local Request Scheduling}
\label{alg:local}
\begin{algorithmic}[1]
\Require A set of $n$ requests $R = \{r_1, r_2, \dots, r_n\}$. Each request $r_i$ has: a prompt length $p_i$, a chunked-prefill speed $c_i$ determined by the model serves it, a TTFT SLO $s_i$, and an arrival time $a_i$.

\Ensure A subset of requests $S \subseteq R$ that can be executed in order to maximize TTFT SLO attainment.

\State Sort $R$ in ascending order of deadlines $d_i = a_i + s_i$: 
$r_1, r_2, \dots, r_n$ such that $d_1 \leq d_2 \leq \dots \leq d_n$.
\State Initialize $S \leftarrow \emptyset$, $current\_time \leftarrow$ Timer.time()

\For{$k=1$ to $n$}
  \State Let $r \leftarrow r_{k}$, $e_r \leftarrow \frac{p_r}{c_r}$
  \State Append $r$ to $S$
  \State Update $current\_time \leftarrow current\_time + e_r$.
  \If{$current\_time > a_{r} + s_{r}$}
    \State \textcolor{blue}{/* pop the request with longest execution time */}
    \State Let $r_{\text{max}} \leftarrow \arg\max\limits_{r' \in S} \frac{p_{r'}}{c_{r'}}$
    \State Remove $r_{\text{max}}$ from $S$
    \State Update $current\_time \leftarrow current\_time - \frac{p_{r_{\text{max}}}}{c_{r_{\text{max}}}}$
  \EndIf
\EndFor

\State \Return $S$

\end{algorithmic}
\end{algorithm}

\MyPara{Analysis.} \tool focuses arbitration on TTFT attainment because it gates when a request can start; once admitted, a request runs through decoding under its TPOT SLO, which is governed by memory headroom rather than queue ordering. Long-running decode requests are preempted only when memory is severely constrained, so a single heavy request cannot stall many newly admitted ones, matching the batch scheduling policies of SGLang and vLLM~\cite{sglang@arxiv23, kwon2023efficient}.

Request scheduling ensures optimal TTFT attainment when chunked-prefill has prefill running at each inference step.
This is because the prefill time $e_r$ of a request $r$ can be estimated as $e_r=\frac{p_r}{c_r}$, allowing us to compute the prefill completion time of any request $r_i$ in a sequence using $d_{r_i}=a_{r_i} + \sum_{i=1}^{n}\frac{p_{r_i}}{c_{r_i}}$, where $n$ is the number of requests (including $r_i$) waiting for processing.
With this information, we can follow the proof of the original Moore-Hodgson algorithm~\cite{cheriyan2021simple} to prove the optimality of our scheduling algorithm.
Our admission control ensures prefill runs at each inference step (so no starvation) by admitting a proper number of requests to each engine and preempting long decoding requests.

%% file: eval-v1.tex
\mysection{Evaluation}
\label{sec:eval}

We implemented a prototype of \tool with $\sim$10,400 lines of Python and 774 lines of C++ code. As the serving backend, we used SGLang~\cite{sglang@arxiv23}, a widely adopted open-source inference engine, and extended it with our elastic memory manager library.
The library is built on CUDA VMM APIs~\cite{cuda-vmm} and exposes standard KV cache allocation APIs via Python bindings, requiring only 22 lines of changes to integrate with SGLang.

On the frontend, we used a Redis queue~\cite{redis-list} to cache incoming requests from all clients. \tool's local scheduler dispatches these requests to the serving engines of corresponding models based on Algorithm~\ref{alg:local}.
For tensor-parallel models, the GPU-local scheduler runs only on the first rank, and the resulting scheduling decisions are broadcast to all other ranks to ensure consistency.
\tool's global scheduler operates as a separate Python process, collecting execution metrics from each engine\textemdash such as request rates and queue status. It makes scheduling decisions (\eg, model evictions and activations) and communicates them to the engines using ZeroMQ~\cite{zmq}.

\mysubsection{Experimental Setup}

\MyPara{Testbed.}
We conducted our experiments on a cluster of four nodes, each equipped with eight NVIDIA H100-80G GPUs interconnected via 600GB/s NVLink. These nodes communicate through a 100Gbps Ethernet network. Each node features two 52-core Intel Xeon Platinum 8480+ CPUs, 1.7 TB of DRAM, and a PCIe Gen5 x16 interface. All nodes run Ubuntu 22.04 with CUDA Toolkit 12.4.

\MyPara{Baselines.}
We compared \tool against four baselines that share GPUs to serve multi-LLMs: (1) \textit{Static partition (S-Partition)}; (2) \textit{MuxServe++}; (3) \textit{QLM~\cite{qlm}}; and (4) ServerlessLLM~\cite{fu2024serverlessllm}.
Note that the original MuxServe is built on vLLM and supports only Llama-2 models. We ported it to SGLang and generalized its memory mechanism with our kvcached to support heterogeneous models, referred to as \muxp.
We evaluated the performance of MuxServe++ and MuxServe using three Llama-3.1-8B models under different request rates over a 10-minute period: 199 requests/min, 262 requests/min, and 22 requests/min. All experiments were conducted under the same and consistent conditions. The results are shown in Table~\ref{tab:muxserve-comparison-slim}.
As we can see, MuxServe++ achieves comparable or even better performance.

\begin{table}[t]
\centering
\begin{adjustbox}{max width=\linewidth}
\begin{tabular}{lcccc}
\toprule
 & Mean E2E (s) & P95 E2E (s) & Req Tput (r/s) & Token Tput (t/s) \\
\midrule
MuxServe    & 7.40 & 18.85 & 7.97 & 3363.53 \\
MuxServe++  & 5.25 & 12.09 & 7.98 & 3353.09 \\
\midrule
 & Mean TTFT (s) & P95 TTFT (s) & Mean TPOT (ms) & P95 TPOT (ms) \\
MuxServe    & 1.47 & 11.04 & 21.21 & 31.95 \\
MuxServe++  & 0.089 & 0.320 & 18.82 & 33.97 \\
\bottomrule
\end{tabular}
\end{adjustbox}
\caption{Performance comparison of MuxServe and MuxServe++}
\label{tab:muxserve-comparison-slim}
\end{table}

Note that general GPU-sharing systems target application-agnostic workloads and cannot be directly compared: they focus on single-GPU sharing rather than cluster-level GPU sharing, and are not designed to optimize LLM serving, which requires both TTFT and TPOT adherence (\S\ref{sec:rw}).

\MyPara{Traces and models.} 
We used two real-world traces, Hyperbolic~\cite{hyperbolic} and Arena-Chat~\cite{chatbotarena@arxiv24}. For each trace, we sampled a representative set of models, including both popular and long-tail models, ensuring their workload characteristics (\eg, popularity distribution and idle patterns) align with the observations in \S\ref{sec:motivation}.
To simulate various scenarios, we scaled the traces by multiplying the number of requests by a factor of $N$, increasing the load while preserving the original traffic patterns---a common way used in prior work~\cite{alpaserve@osdi23, muxserve}.
In total, we evaluated 58 LLMs as detailed in Table~\ref{tab:models}. The large-scale experiments (\S\ref{sec:large-scale}) use all models, while other evaluations (\S\ref{sec:e2e}---\S\ref{sec:ablation}) select subsets tailored to specific goals.

\begin{table}[t]
\centering
\footnotesize
\begin{tabular}{lcccc}
\toprule
\textbf{Model size} & 1B-3B & 4B-8B & 9B-30B & 31B-70B \\
\midrule
\#LLMs & 43 & 8 & 3 & 4 \\
\bottomrule
\end{tabular}
\vspace{-2mm}
\caption{Models used in our evaluation.}
\vspace{-1mm}
\label{tab:models}
\end{table}

\begin{figure*}[th]
  \centering
  \includegraphics[width=0.8\textwidth]{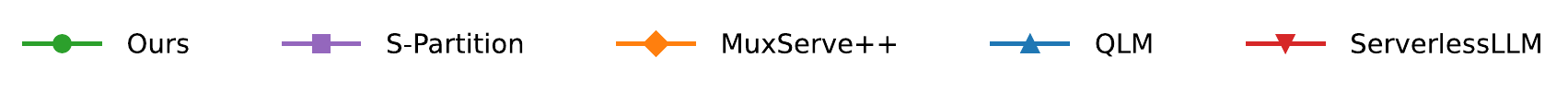}
  \vspace{-2mm}
  
  \begin{subfigure}[t]{0.44\textwidth}
    \includegraphics[width=\linewidth]{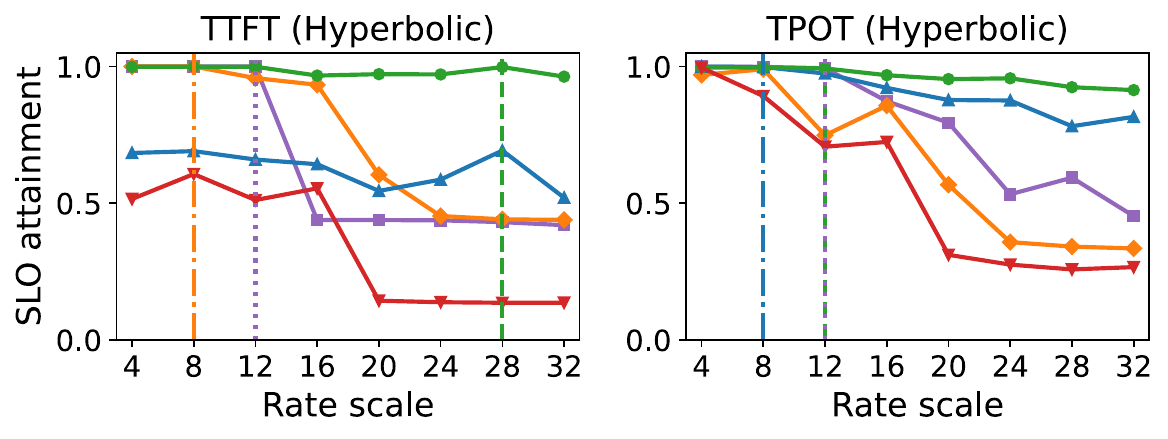}
  \end{subfigure}
  \begin{subfigure}[t]{0.44\textwidth}
    \includegraphics[width=\linewidth]{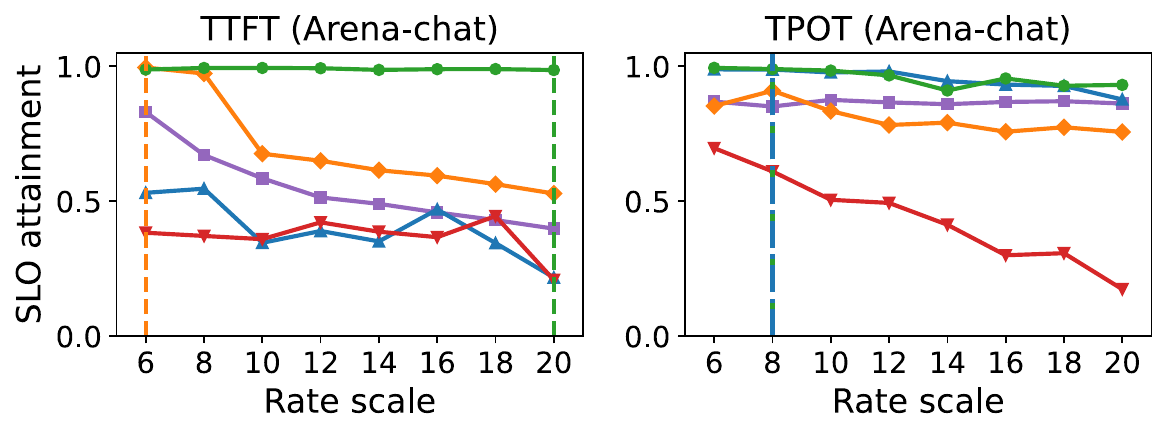}
  \end{subfigure}
  \vspace{-1.8mm}

  \begin{subfigure}[t]{0.44\textwidth}
    \includegraphics[width=\linewidth]{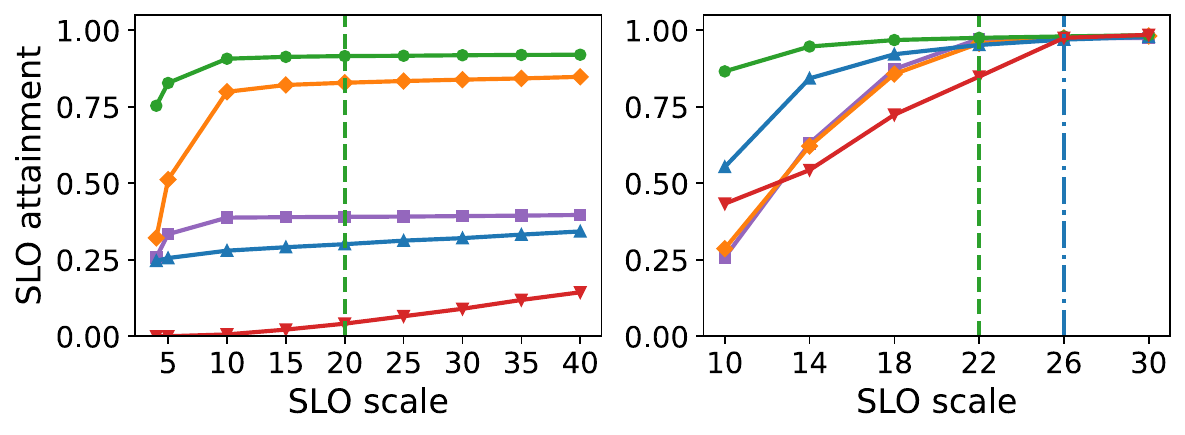}
  \end{subfigure}
  \begin{subfigure}[t]{0.44\textwidth}
    \includegraphics[width=\linewidth]{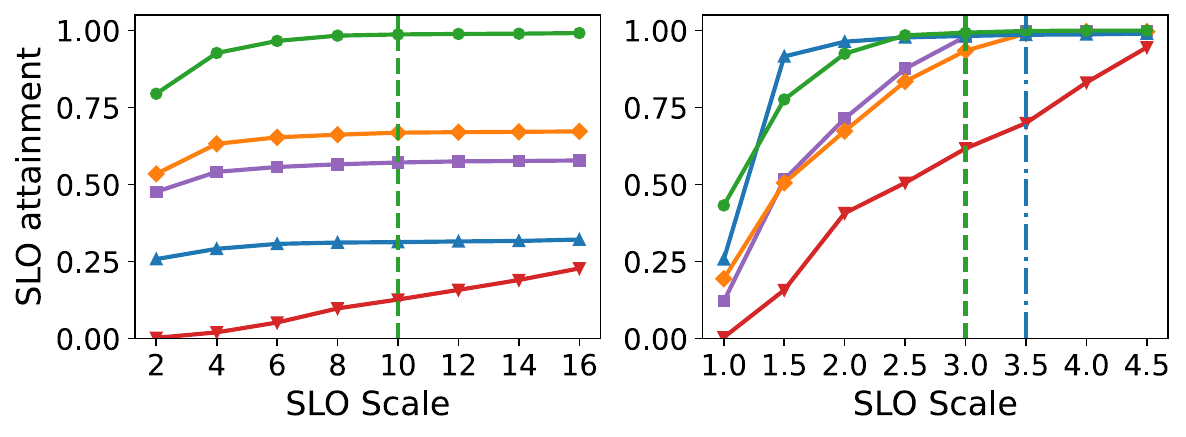}
  \end{subfigure}
  \vspace{-1.8mm}

  \begin{subfigure}[t]{0.44\textwidth}
    \includegraphics[width=\linewidth]{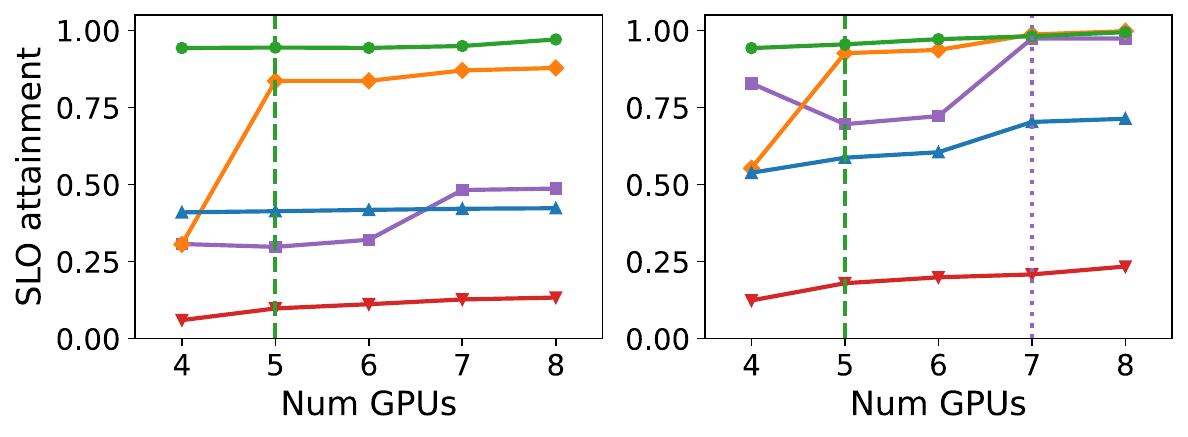}
  \end{subfigure}
  \begin{subfigure}[t]{0.44\textwidth}
    \includegraphics[width=\linewidth]{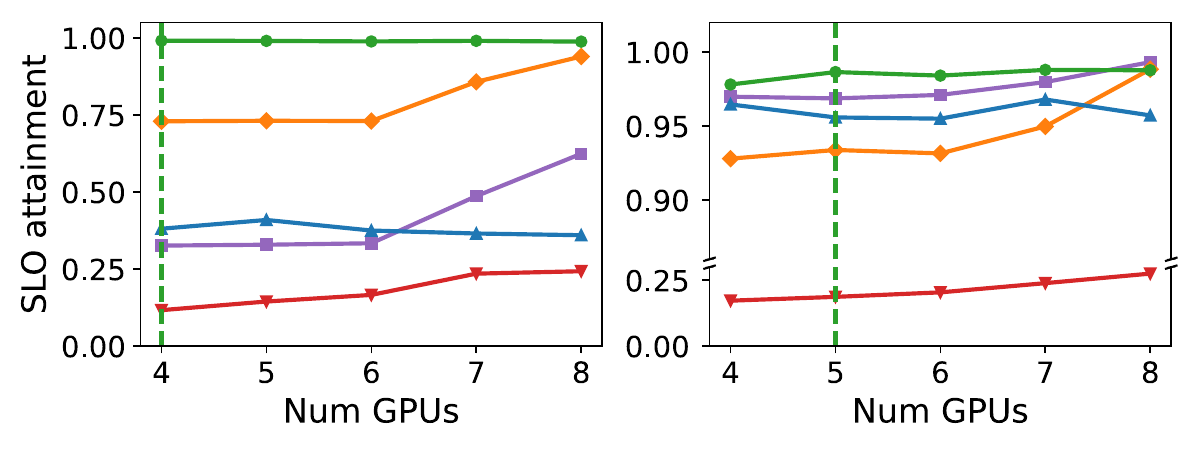}
  \end{subfigure}
  \vspace{-3mm}
  \caption{End-to-end performance comparison on SLO attainments under varying scales of rates, SLOs, and numbers of available GPUs. The dotted vertical lines mark where the system reaches 99\% TTFT or TPOT attainment.}
  \label{fig:slo-comparison}
  \vspace{-1em}
\end{figure*}
\MyPara{Metrics.} 
Our primary performance metrics are TTFT and TPOT attainment.
To establish SLOs for each model, we first ran its workload on dedicated GPUs to measure its P95 TTFT and TPOT latencies. This process produced TTFT SLOs ranging from 0.04s to 0.13s and TPOT SLOs from 5.2ms to 50.9ms.
We then scaled these base values by a factor to evaluate system performance under varying latency requirements, following an approach consistent with prior work~\cite{alpaserve@osdi23, muxserve, qin2025mooncake}.
We also reported aggregated throughput. To account for model idle periods, throughput considers the actual time (excluding idle time) spent serving them.

\mysubsection{End-to-End Performance} 
\label{sec:e2e}

We start with the end-to-end performance of \tool under varying request rates, SLO requirements, and GPU supplies.

\MyPara{SLO attainment vs. request rate.}
We first evaluated various inference loads by serving eight models on two shared GPUs.
As shown in Figure~\ref{fig:slo-comparison} (first row), \tool consistently outperforms all baselines by maintaining a significantly higher TTFT SLO attainment.
On the Hyperbolic trace, \tool supports up to $2.3\times$ and $3.5\times$ more requests than \muxp and static partitioning, respectively, while still achieving 99\% SLO attainment. On the Arena-Chat trace, it handles over $3\times$ more requests than all baselines. The gains stem from KVPR-driven placement, which keeps per-GPU memory pressure low so active models retain headroom under bursty arrivals.
\muxp's TTFT SLO attainment drops quickly with higher request rates because it cannot evict idle models or relocate models across GPUs, leading to memory contention and degraded performance.

QLM time-shares GPUs by packing pending requests into groups and dispatching each group to a GPU under an SLO-aware stochastic policy, swapping models and preempting unfinished requests whenever consecutive groups target different models. The frequent swapping makes QLM's TTFT attainment worse than static partitioning.

ServerlessLLM serves models in a time-sharing manner for serverless workloads: inactive models are unloaded, and newly arriving requests wait until the scheduler reactivates the required model on the server with the fastest startup, selected based on checkpoint locality. Because the full cold-start process still dominates request latency, ServerlessLLM achieves the worst TTFT SLO attainment.

\tool maintains high TPOT attainment through its demand-aware scheduler, which balances workloads to reduce memory contention. Although the scheduler explicitly targets TTFT, TPOT benefits as a side effect: TPOT is degraded mainly by oversized batches and memory-induced preemptions, both of which subside once KVPR-balanced placement caps per-GPU memory contention. QLM achieves lower TPOT because it over-batches under high load, increasing per-iteration latency, while ServerlessLLM performs even worse by allowing unbounded batch sizes that further raise P99 TPOT. Both \muxp and static partitioning experience severe memory contention at high request rates, causing frequent preemptions that substantially degrade TPOT.

\MyPara{SLO attainment vs. SLO requirements.} 
The middle row of Figure~\ref{fig:slo-comparison} shows the attainment across different SLO targets.
As SLO scales up, \tool quickly achieves 99\% TTFT and TPOT attainment. 
In contrast, no baseline achieves 99\% TTFT attainment even on the largest SLO scale in this experiment, and their attainment rates do not improve significantly as the scale increases. 
Among the baselines, \muxp achieves the best TTFT performance, reaching 84.79\% and 67.22\% attainment on the largest SLO scale for the two traces. The gap to \tool is primarily due to their inflexibility to adapt sharing policies dynamically. 
The TPOT attainment of all systems increases rapidly as the SLO scale grows 
because TPOT is less sensitive to memory contention. 
We also observe that the Hyperbolic trace requires higher SLO scales due to more bursty and heavier request patterns.

\MyPara{SLO attainment vs. available GPUs.}
Finally, we evaluate performance when provisioning more GPUs. We selected 18 models from Table~\ref{tab:models}, representing a mix of popular and tail models with diverse load variability.
To fully test the flexibility of our scheduling strategy, we included models of varying sizes from 1B to 8B, all of which fit within a single 80GB GPU. This setup enables a wide range of model-sharing combinations across GPUs.
As shown in Figure~\ref{fig:slo-comparison} (last row), \tool achieves 99\% TTFT and TPOT attainment using only four and five GPUs on the two traces, respectively, demonstrating its effectiveness in improving cost-efficiency while maintaining performance.
In comparison, all baselines fail to reach 99\% TTFT attainment even with eight GPUs, and only a few baselines achieve 99\% TPOT attainment when seven or eight GPUs are provisioned.

\begin{figure}[t]
    \centering
    \begin{subfigure}[b]{\linewidth}
        \centering
        \includegraphics[width=0.82\linewidth]{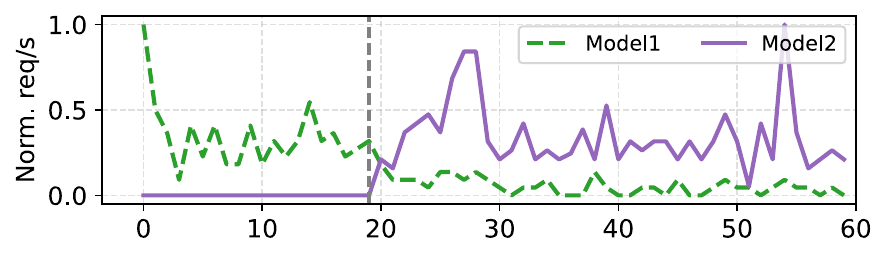}
    \end{subfigure}

    \begin{subfigure}[b]{\linewidth}
        \centering
        \includegraphics[width=0.82\linewidth]{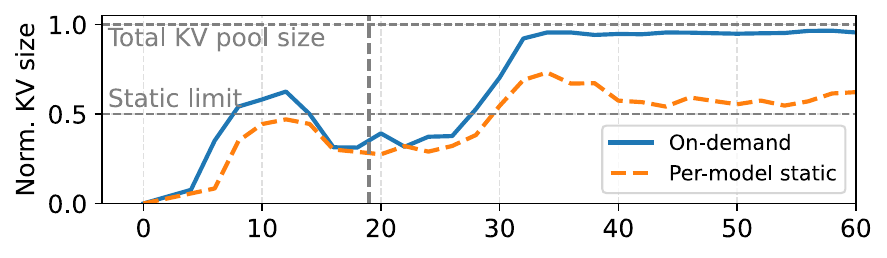}
    \end{subfigure}

    \begin{subfigure}[b]{\linewidth}
        \centering
        \includegraphics[width=0.82\linewidth]{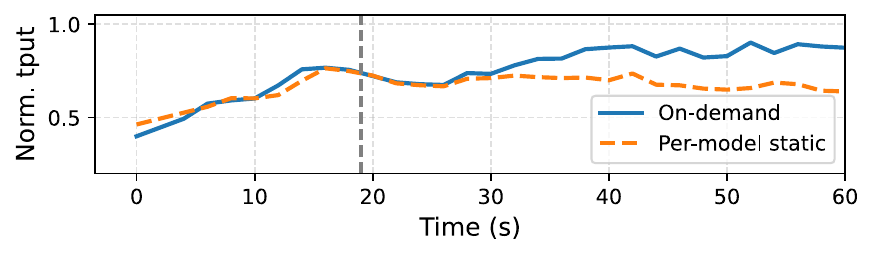}
    \end{subfigure}
   \vspace{-6mm}
    \caption{Benefits of cross-model memory coordination. The first figure shows the request rates. The last two figures shows the total KV memory size and the throughout of the two models.}
    \label{fig:elastic-mem-microbench}
\end{figure}

\mysubsection{Performance Analysis}
\label{sec:ablation}

Next, we provide a detailed performance breakdown to analyze the effectiveness of each design in \tool and how each contributes to its strong end-to-end performance.

\MyPara{Flexible cross-model memory sharing.}
We first evaluated the benefits of our flexible cross-model memory sharing by comparing with static partition using a simplified two-model trace extracted from Arena-Chat, shown in Figure~\ref{fig:elastic-mem-microbench} (first row).
We present the normalized total KV cache usage and aggregated throughput for both methods in the last two rows of Figure~\ref{fig:elastic-mem-microbench}.
As we can see, \tool's on-demand memory allocation allows it to use more memory for KV cache, particularly after the 20th second, when Model1 experiences low demand while Model2 faces a surge in request rates.
The larger KV cache memory enables \tool to achieve higher throughput, as shown in the last row. 
In contrast, under static partitioning, even when Model1 underutilizes its memory, Model2 cannot leverage the unused memory due to the static allocation boundary. 

\begin{figure}[t]
  \centering
  \begin{subfigure}[t]{0.46\linewidth}
    \includegraphics[width=\linewidth]{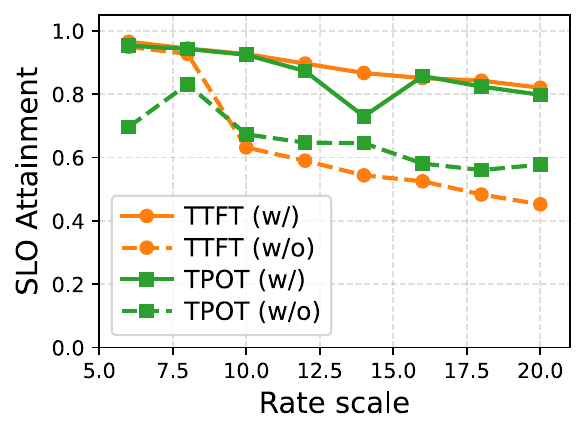}
    \vspace{-6mm}
    \caption{Attainment with rate scales}
    \label{fig:global-scheduler-attainment}
  \end{subfigure}
  \begin{subfigure}[t]{0.49\linewidth}
    \includegraphics[width=\linewidth]{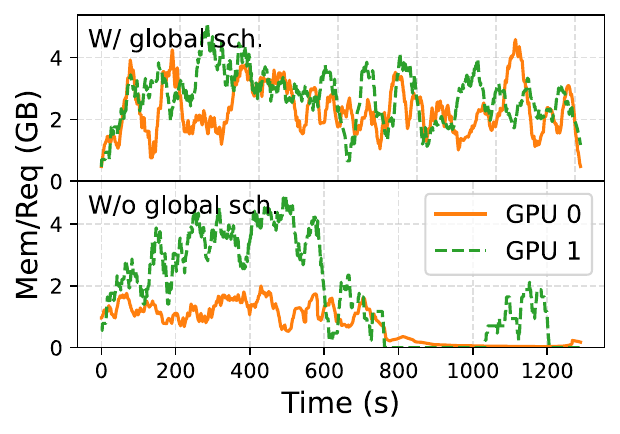}
    \vspace{-6mm}
    \caption{GPU load status}
    \label{fig:global-scheduler-load}
  \end{subfigure}
  \caption{Effectiveness of global model placement scheduling.}
  \label{fig:global-scheduler}
\end{figure}

\MyPara{Model placement.}
Next, we evaluated the benefits of our global scheduler that conducts model placement.
In this experiment, we used two GPUs to serve eight models from Arena-Chat. Figure~\ref{fig:global-scheduler-attainment} presents the TTFT and TPOT attainment with the global scheduler enabled or disabled.
The results show that enabling the global scheduler significantly improves both TTFT and TPOT attainment.
To provide further insights, we plot the average KV cache memory available per request as it arrives on each GPU. With the global scheduler enabled, the load is more evenly balanced across the two GPUs, allowing each request to access more KV cache memory on average.
In contrast, without the global placement scheduler, the load is imbalanced: GPU1 shows more available memory during the first 600 seconds, while the near-zero availability between the 800th and 1000th seconds indicates that GPU1 is idle while GPU0 is overloaded.

\begin{figure}[t]
  \centering
  \begin{subfigure}[t]{0.48\linewidth}
    \includegraphics[width=\linewidth]{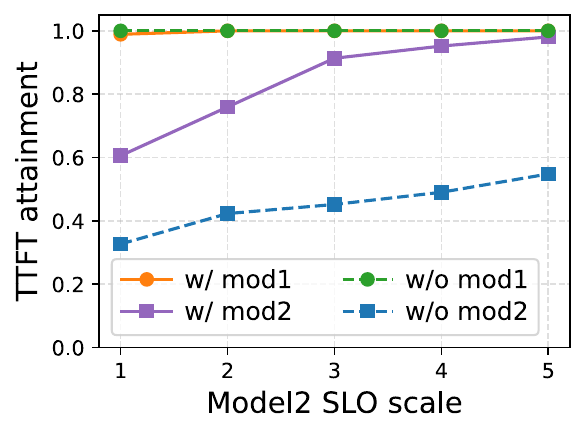}
    \caption{Attainment with SLO scales}
    \label{fig:gpu-scheduler-slo}
  \end{subfigure}
  \begin{subfigure}[t]{0.48\linewidth}
    \includegraphics[width=\linewidth]{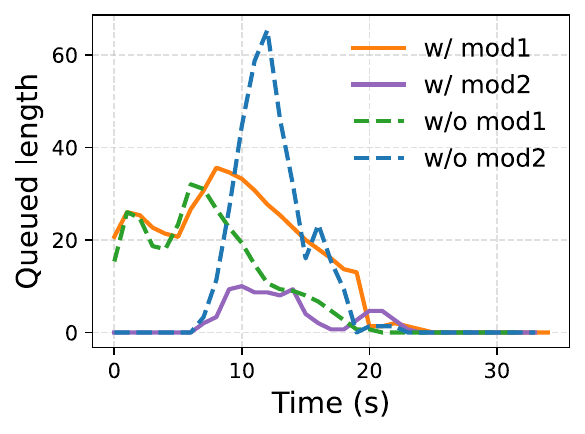}
    \caption{Queue length with time}
    \label{fig:gpu-scheduler-qlen}
  \end{subfigure}
  \caption{Effectiveness of GPU local request scheduling.}
  \label{fig:gpu-scheduler}
\end{figure}

\MyPara{Request arbitration.}
To evaluate the GPU-local scheduler that prioritizes requests, we use two models: we fix the SLO scale of Model1 to eight and vary the SLO scales of Model2 to evaluate the priority-based admission control in the GPU-local scheduling.
Figure~\ref{fig:gpu-scheduler-slo} shows the TTFT attainment as we vary the SLO scale of Model2. Model1 consistently maintains high attainment, and enabling our GPU-local scheduling improves the SLO attainment of Model2 by more than 40\%.
To dive deeper, we plot the queue length of each model in Figure~\ref{fig:gpu-scheduler-qlen} of one experiment run.
From the queue lengths, we clearly observe that when the local scheduler is enabled, the system prioritizes Model2's requests, which are shorter but have stricter SLO requirements. 

\mysubsection{Large-Scale Evaluation}
\label{sec:large-scale}

Finally, we evaluated how effectively \tool in reducing the cost of multi-LLM serving at scale.
We served all 58 models listed in Table~\ref{tab:models}, following common TP practices for large models: TP=4 for 32B models~\cite{qwen2.5, qwen2.5hf} and TP=4 or 8 for 70B models~\cite{llama3, llama3.3hf}, utilizing 32 GPUs in total.
We sampled 58 models from the Arena-Chat trace and the Hyperbolic trace.

\begin{figure}[t]
  \centering
  \begin{subfigure}[t]{0.98\linewidth}
    \includegraphics[width=\linewidth]{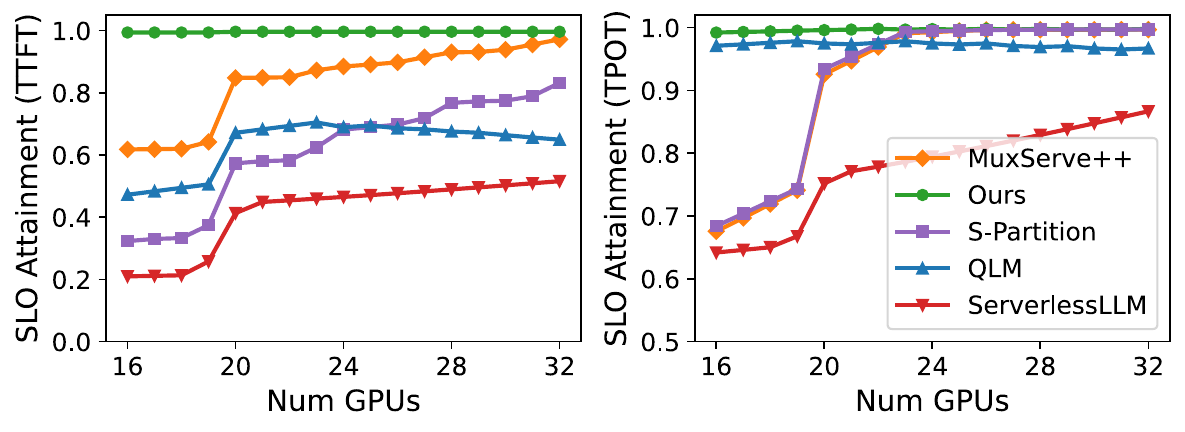}
   \vspace{-6mm}
    \caption{SLO attainment with cluster size scaling}
    \label{fig:large-scale-attainment}
  \end{subfigure}
  \begin{subfigure}[t]{0.98\linewidth}
    \includegraphics[width=\linewidth]{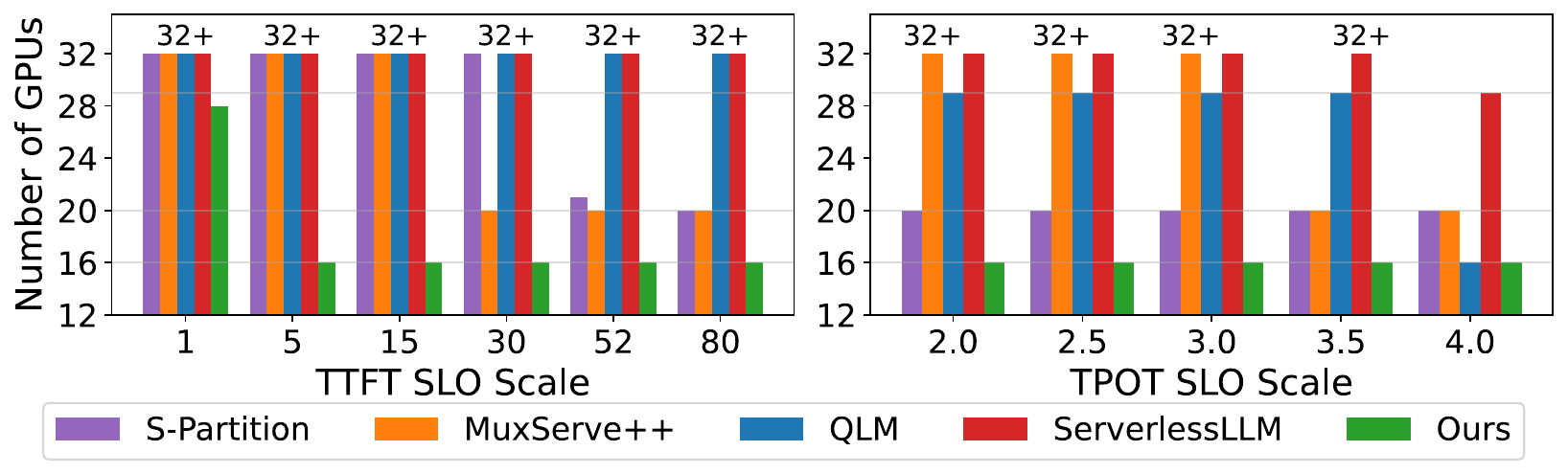}
   \vspace{-4mm}
    \caption{Number of GPUs needed for 99\% SLO attainments}
    \label{fig:large-scale-gpus}
  \end{subfigure}
  \vspace{-2mm}
  \caption{SLO attainment and cost saving at large scales.}
  \label{fig:large-scale}
  \vspace{-2mm}
\end{figure}

\MyPara{SLO attainments vs. number of GPUs.} 
Figure~\ref{fig:large-scale-attainment} shows TTFT and TPOT attainment with an increased number of GPUs.
\tool outperforms all baselines, achieving nearly 99\% TTFT attainment with just 16 GPUs, while \muxp requires 32 GPUs to reach similar performance, and other methods require even more.
As the number of GPUs increases, all baselines improve in TTFT and TPOT attainment except QLM.
We find this to be related to its suboptimal scheduling algorithm.
QLM assigns requests to GPUs without considering which models are already on GPUs; when a queue is empty, it picks the first available GPU, often triggering unnecessary swaps. With more GPUs, this leads to more idle devices and frequent swapping. Its inefficient swapping, which requires engine restart~\cite{qlm2024github}, adds significant latency, causing queued requests to miss their SLOs.

\begin{figure}[t]
    \centering
    \includegraphics[width=0.90\linewidth]{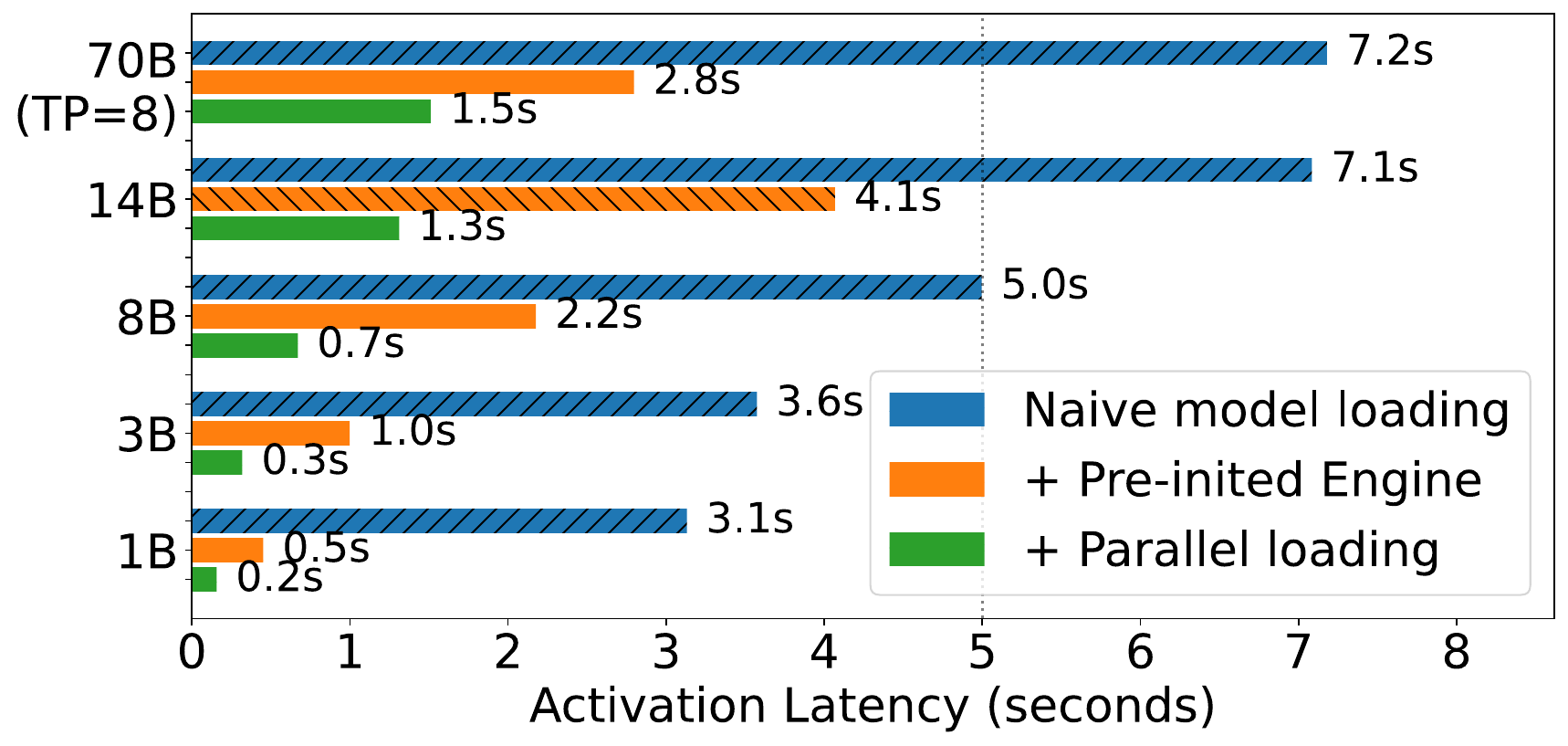}
    \vspace{-3mm}
    \caption{The activation time for models with different sizes. Data is measured on H100 GPUs.}
    \vspace{-2mm}
    \label{fig:loading-time}
\end{figure}

\MyPara{Cost saving.} 
Figure~\ref{fig:large-scale-gpus} shows the number of GPUs required to achieve 99\% SLO attainments at different SLO scales.
If a system fails to achieve 99\% attainment with all 32 GPUs, we denote its GPU requirement as ``32+''. 
\tool achieves 99\% TTFT SLO attainments with only 16 GPUs when SLO scale is 5 and TPOT SLO scale is 2.0. 
\muxp needs 20 GPUs to get 99\% TTFT SLO attainments with SLO scale $\geq$ 30, while static partition needs even more GPUs or a higher SLO scale.
For TPOT, static partition is the best across all baselines, requiring 20 GPUs, while QLM and \muxp need at least 29 GPUs when TPOT SLO scale $\leq$ 3.0.

\begin{figure}[t]
  \centering
  \begin{subfigure}[t]{0.48\linewidth}
    \includegraphics[width=\linewidth]{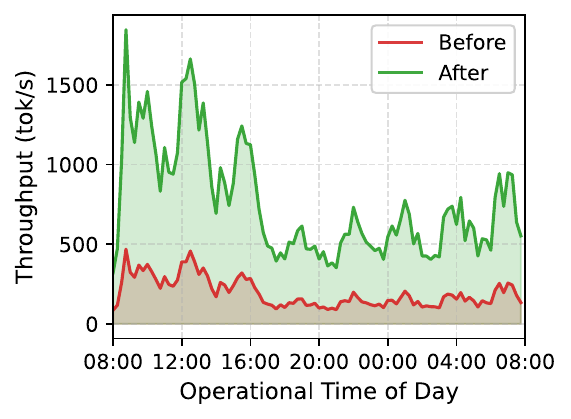}
    \caption{Company A}
    \label{fig:production-throughput-a}
  \end{subfigure}
  \begin{subfigure}[t]{0.48\linewidth}
    \includegraphics[width=\linewidth]{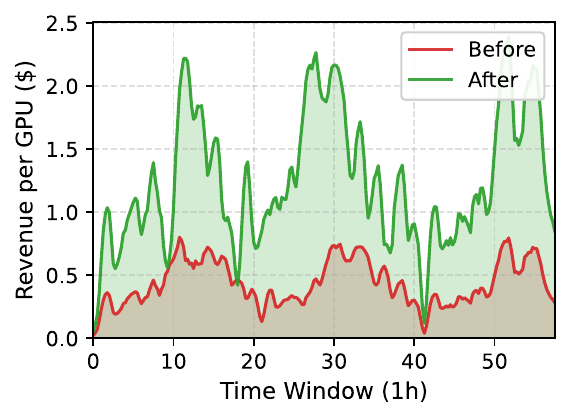}
    \caption{Company B}
    \label{fig:production-throughput-b}
  \end{subfigure}
  \caption{Production results from two companies, showing (a) throughput and (b) revenue per GPU before and after using \tool.}
  \label{fig:production-throughput}
\end{figure}

\mysubsection{System Overhead}

\MyPara{Model activation latency.}
We also measured model activation latency from pageable CPU memory for models ranging from 1B to 70B parameters (see \S\ref{sec:model-loading} for more details).
Our optimizations significantly reduce activation latency. \tool loads small models (1B$\sim$8B) within 0.7s, a medium-size model (14B) in just 1.3s, and large models (>70B) in 1.5s (Figure~\ref{fig:loading-time}).
These results show that \tool can promptly activate evicted models upon receiving new requests.

\MyPara{Activation and migration frequency.}
Over a 10-minute window of the eight-models-on-two-GPUs run, \tool issued 2 idle-driven activations and 3 inter-GPU migrations. Migrations stay off the TTFT critical path: weights and in-flight KV state transfer over NVLink within tens of milliseconds (\eg, $\sim$20\,ms for an 8B model) while the source instance keeps serving. Activations complete within 2\,s, well inside the multi-second TTFT SLOs typical of online serving.

\MyPara{Elastic memory overhead.}
We evaluated \tool's elastic memory overhead in the worst case, where the request rates are constant, leaving no opportunities for dynamic sharing. With two colocated Llama-3.2-3B models on an A100-40G GPU, \tool incurs
 only 3 ms (4 \%) TTFT and 4 ms (13 \%) TPOT overhead at a high load of 32 req/s and
only 2 ms (3.5 \%) TTFT and 3 ms (7 \%) TPOT overhead at a high load of 28 req/s (see~\S\ref{sec:elastic-overhead}), compared to static partitioning.

\mysubsection{Production Workloads}
As of December 2025, \tool has been deployed in multiple organizations serving LLM workloads. Figure~\ref{fig:production-throughput} summarizes deployments at a large tech company (Company A) using \tool on an internal GPU cluster, and a commercial LLM inference provider (Company B) serving external customers.
To isolate \tool's contribution from workload variation, both deployments use shadow workload replay: the same online request stream is mirrored to two identical clusters, one running \tool and the other not, so that the model mix and traffic patterns are held constant across the two arms.

Figure~\ref{fig:production-throughput-a} shows Company A’s per-GPU token throughput before and after adopting \tool. Company A serves diverse applications using fine-tuned and off-the-shelf models (3B–70B) on a shared GPU pool. With \tool dynamically adjusting model placement and sharing\textemdash without changes to existing inference engines\textemdash Company A achieves consistently higher throughput (3.89× on average) over several weeks, with no SLO violations and unchanged tail latency. Figure~\ref{fig:production-throughput-b} presents results from Company B, which optimizes revenue per GPU under highly variable traffic, computed as the prefill and decode tokens generated in the window priced at the provider's published per-token rates, normalized by GPU count. Prior to \tool, fragmentation and overprovisioning constrained revenue; after deployment, \tool improves utilization while meeting all SLOs, raising revenue per GPU by 2.86$\times$ by converting idle capacity into billable tokens.

%% file: rw-v2.tex
\mysection{Related Work}
\label{sec:rw}

\noindent\textbf{SLO-aware LLM scheduling.}
Llumnix~\cite{sun2024llumnix}, SLOs-Serve~\cite{chen2025slos}, ExeGPT~\cite{oh2024exegpt}, SAGESERVE~\cite{sageserve}, DistServe~\cite{zhong2024distserve}, and MELL~\cite{qianli2025mell} explore SLO-aware scheduling to improve LLM inference performance.
However, they primarily focus on request scheduling or resource allocation for single-model serving, while \tool enables dynamic cross-model memory harvesting for efficient multi-model serving.

\noindent\textbf{Memory management in LLM serving.} 
vAttention~\cite{prabhu2024vattentiondynamicmemorymanagement} and vTensor~\cite{vtensor} use CUDA virtual memory APIs~\cite{cuda-vmm} to decouple physical and virtual memory allocation.
However, their purpose is to simplify programming 
and improve kernel efficiency for single-LLM serving. 
In addition, they require re-implementing a large portion of the current inference engine stack, while our method preserves compatibility with the widely used PagedAttention~\cite{kwon2023efficient} mechanism.

\noindent\textbf{GPU sharing techniques.}
GPU sharing has been extensively studied~\cite{zhang2025sgdrc, zhang2025improving, ng2023paella, strati2024orion, han2022microsecond, han2024pantheon, tgs, yu2019salus, fgd@atc23, antman@osdi20, reef@osdi22, IADeep@sc23, zhao2023muxflowefficientsafegpu, chen2017effisha, wang2024gcaps, alpaserve@osdi23, muxserve, fu2024serverlessllm, qlm, aegaeon}, but largely at the OS or runtime layer for application-agnostic, single-GPU workloads. Compute-centric designs time-slice between latency-critical and best-effort co-tenants~\cite{tgs, gpreempt@atc25} or add preemptive scheduling across heterogeneous accelerators~\cite{xsched@osdi25}, whereas \tool targets memory sharing across uniformly latency-critical LLMs spanning multiple GPUs. The closest memory-centric work, MSched~\cite{msched}, schedules GPU\textendash CPU paging \emph{after} oversubscription, while \tool proactively prevents oversubscription via coordinated placement.  Existing multi-LLM solutions remain fragmented: MuxServe~\cite{muxserve} supports spatial sharing but pins models to devices even when idle, while Aegaeon~\cite{aegaeon} performs temporal sharing via prefill–decode disaggregation, requiring weight duplication and struggling with diverse, dynamic access patterns. \tool unifies spatial and temporal sharing through elastic GPU memory management.

%% file: conclusion.tex
\mysection{Conclusion}

This paper introduces \tool, a multi-LLM serving system that improves cost efficiency while maximizing SLO attainment via GPU sharing. 
\tool achieves this by enabling flexible memory ballooning, 
and employing a two-level scheduling algorithm to make efficient use of GPU memory. 

\vspace{-.5em}
\section*{Acknowledgments}
\vspace{-.5em}
We thank the anonymous reviewers for their comments and are particularly grateful to our shepherd Xiaosong Ma for her valuable feedback.
We also thank Ion Stoica, Matei Zaharia, Jiangfei Duan, Chenxi Wang, Shi Liu, Zhenting Zhu, and Yicheng Liu for their discussions and insights.
We are grateful to Hyperbolic, Novita AI, Chatbot Arena, NVIDIA, and the SGLang open-source community for production traces and hardware support.
This work was supported by an Amazon AI Fellowship (awarded to Shan Yu), the National Science Foundation (CNS-1763172, CNS-2007737, CNS-2006437, CNS-2106838, CNS-2147909, CNS-2128653, CNS-2301343, CNS-2330831, CNS-2403254, IIS-2546642), the Alibaba Research Intern Program, and gifts to Sky Computing Lab from industry partners (Accenture, AMD, Anyscale, Cisco, Google, IBM, Intel, Intesa Sanpaolo, Lambda, Lightspeed, Mibura, Microsoft, NVIDIA, Samsung SDS, and SAP). Junyi Shu, Jiarong Xing, Harry Xu, and Ying Sheng are the corresponding authors.

%% file: appendix.tex
\section{Appendix}
\label{sec:appendix}

\subsection{Analysis of All production traces}
\label{ap:trace}

\begin{figure}[t]
  \centering
  \begin{subfigure}[t]{0.48\linewidth}
    \includegraphics[width=\linewidth]{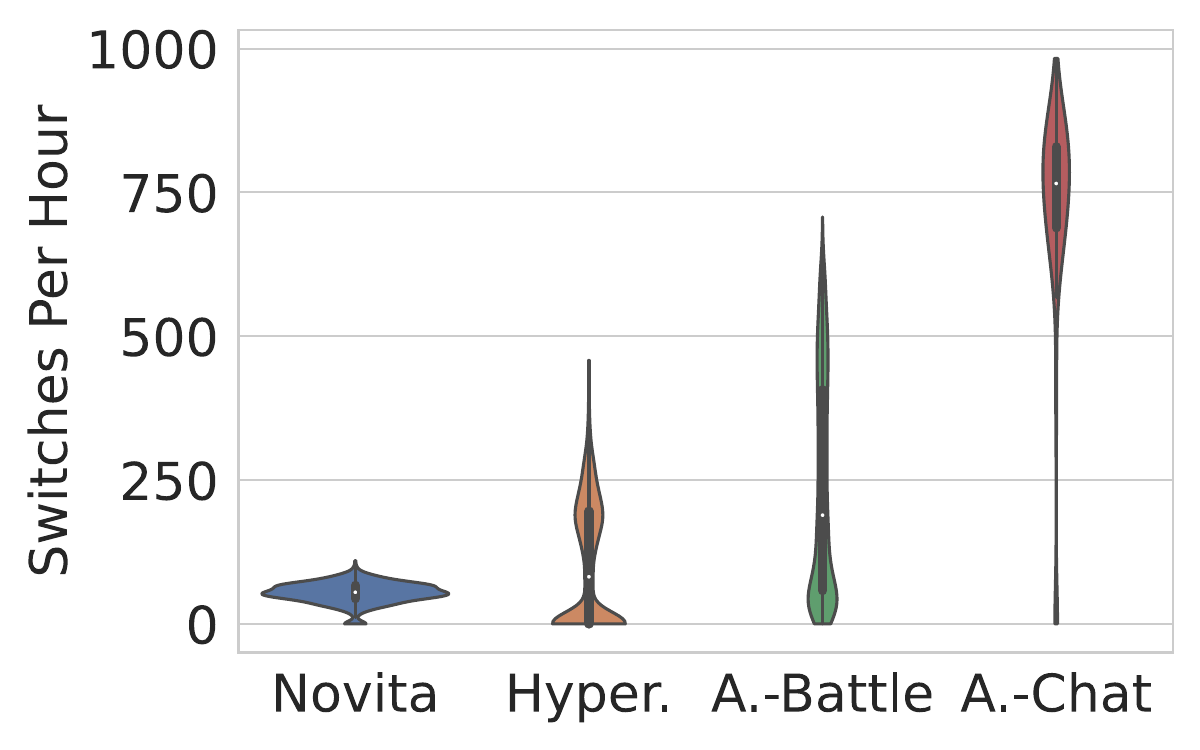}
    \vspace{-5mm}
    \caption{Bursty-group shifts.}
    \label{fig:bursty_shifts}
  \end{subfigure}
  \begin{subfigure}[t]{0.48\linewidth}
    \includegraphics[width=\linewidth]{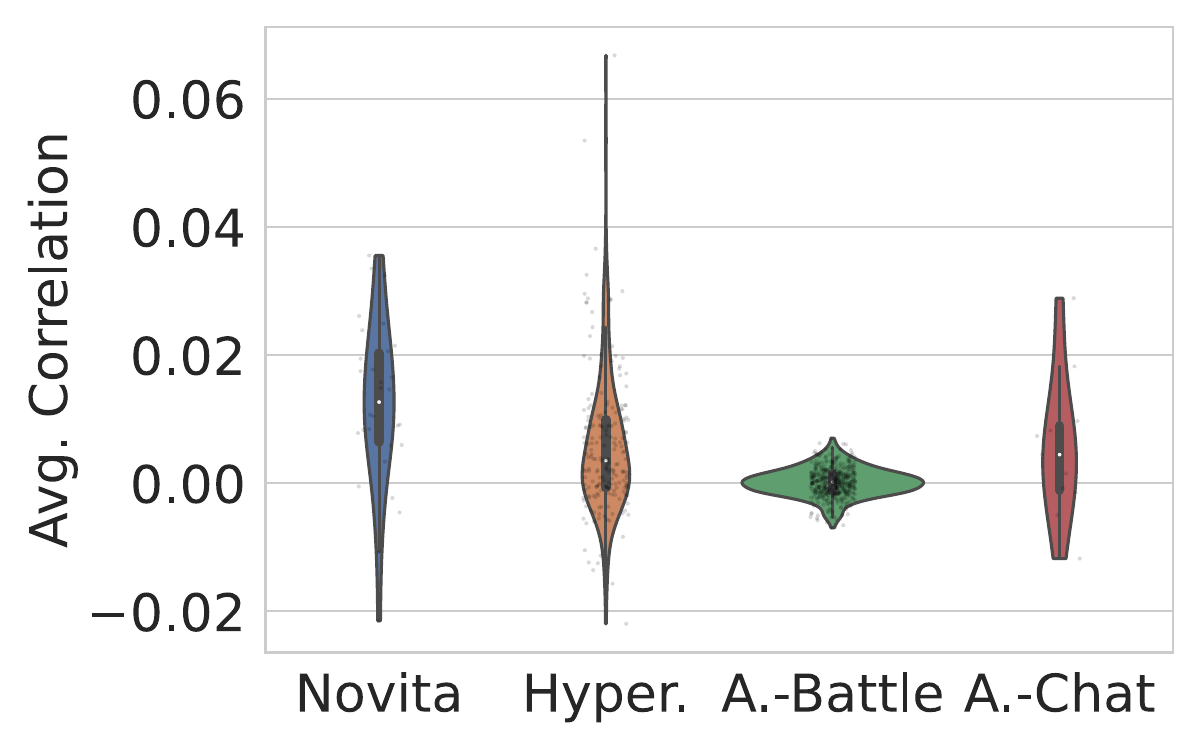}
    \vspace{-5mm}
    \caption{Request rate correlation.}
    \label{fig:corr1}
  \end{subfigure}
  \vspace{-2mm}
  \caption{Request patten shifts.}
  \vspace{1mm}
  \label{fig:corr}
\end{figure}

\MyPara{Overview.}
To understand the workload of multi-LLM serving, we analyze four production traces in detail. These traces are collected from representative service providers as summarized in Table~\ref{tab:trace_summary}.
The first two traces are from Hyperbolic~\cite{hyperbolic} and Novita AI~\cite{novita-ai}, two popular LLM inference service providers. They offer inference APIs for a variety of foundation models and also support user-deployed, fine-tuned models. 
The last two traces are from Chatbot Arena~\cite{chatbotarena@arxiv24}, a widely used open-source platform for LLM evaluation. It compares model responses via human preference voting (Arena-Battle) and also provides interfaces for real-time conversations with various models (Arena-Chat).

\MyPara{Bursty-group behaviors.}
The bursty groups in the production traces shift rapidly over time, and this behavior can be quantified through model switches. As shown in Figure~\ref{fig:bursty_shifts}, we plot the number of model switches per hour for each trace. To compute this metric, we treat a model as active if it has received at least one request in the past two minutes. A model switch is counted whenever the set of active models changes. Across the four traces, Novita exhibits the fewest switches, yet still averages 54 switches per hour, meaning the active set changes almost once per minute. In contrast, Arena-Chat, which contains the largest and most diverse collection of models, experiences an average of 766 switches per hour, reflecting that its active set changes every few seconds.

\MyPara{Unpredictability.}
The request rate at a given time is also highly unpredictable in our traces. Figure~\ref{fig:corr1} quantifies predictability by measuring, for each model, the Pearson correlation between its request-rate time series on consecutive days. Across all traces, the correlations cluster near zero, showing that a model’s traffic at a given time of day provides virtually no information about its traffic at the same time on the next day.

\begin{figure}[t]
  \centering
  \begin{subfigure}[t]{0.48\linewidth}
    \includegraphics[width=\linewidth]{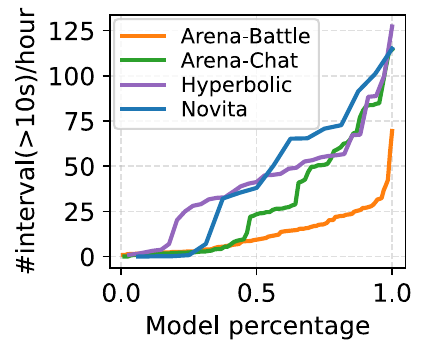}
    \vspace{-5mm}
    \caption{Number of idle intervals per hour.}
    \label{fig:hour-idle-time}
  \end{subfigure}
  \begin{subfigure}[t]{0.48\linewidth}
    \includegraphics[width=\linewidth]{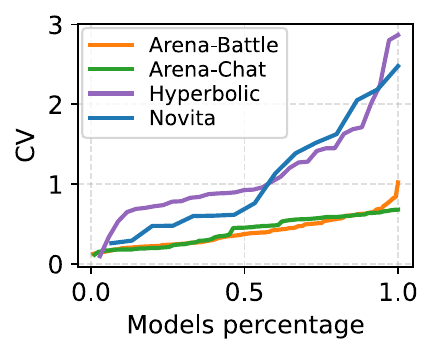}
    \vspace{-5mm}
    \caption{CV of request rate per minute.}
    \label{fig:cv-cdf}
  \end{subfigure}
  \vspace{-2mm}
  \caption{Volatile requests.}
  \vspace{1mm}
  \label{fig:volatile}
\end{figure}

\MyPara{Volatile requests.}
Figure~\ref{fig:hour-idle-time} further shows that models frequently alternate between activity and idleness: many models in the Hyperbolic and Novita traces experience more than 40–100 idle intervals per hour (>10 s each), making static reservations wasteful. Complementing this, Figure~\ref{fig:cv-cdf} reports the coefficient of variation (CV), calculated as the standard deviation divided by the mean ($\sigma/\mu$), of requests per minute for each model. Both Hyperbolic and Novita traces contain many models with CV > 1. The Chatbot Arena traces have lower request rates, resulting in smaller CVs overall, but many models still exhibit CVs greater than 0.5. Together, the high frequency of idle intervals and the substantial variability captured by CV indicate that all traces exhibit strong, persistent volatility.

\subsection{Analysis of Algorithm~\ref{alg:global}}
\label{ap:alg-gloabl}

\subsubsection{KVPR Bound Analysis}
\label{ap:kvpr-bound}

The global model placement algorithm ensures that the maximum KV pressure ratio (KVPR) across all GPUs is bounded by the maximum KVPR in the optimal placement. We give the following analysis.

Let $KVPR_{OPT}$ be the minimum possible maximum KVPR achievable by any optimal placement. Let $KVPR_{max}$ be the maximum KVPR produced by Algorithm \ref{alg:global}. We want to show $KVPR_{max}$ is bounded by $KVPR_{OPT}$.

\MyPara{Bottleneck Analysis:} Focus on the GPU, denoted as $g_{max}$, that achieves the highest KVPR ($KVPR_{max}$) given by Algorithm \ref{alg:global}'s placement. 
Let $m_k$ be the \emph{last} model assigned to this GPU $g_{max}$. Let $W_{before}$ and $S_{before}$ represent the total SLO-weighted request rate and shared KV memory on $g_{max}$ \emph{just before} model $m_k$ was assigned. The final state on this GPU is $KVPR_{max} = (W_{before} +d_k) / (S_{before} - w_k)$, where $d_k$ is the SLO-weighted request rate ($r_k/s_k$) and $w_k$ is the memory weight of model $m_k$.
Similar to Graham~\cite{graham1969bounds}, this proof aims to demonstrate that both the state \emph{before} $m_k$ was added and the contribution \emph{of} $m_k$ are bounded relative to $KVPR_{OPT}$. Specifically, it seeks to establish two conceptual bounds:
\begin{itemize}
    \item \textbf{Bound 1 (Related to state before $m_k$):} The KVPR on $g_{max}$ just before $m_k$'s assignment, $\frac{W_{before}}{S_{before}}$, was the minimum among all GPUs at that moment due to the algorithm's greedy choice. This minimum KVPR is typically related to the average ``pressure'' across the system, which, in turn, is argued to be no larger than the optimal maximum pressure. This suggests the inequality: $W_{before} / S_{before} \le KVPR_{OPT}$
    \item \textbf{Bound 2 (Related to model $m_k$):} The ``pressure'' exerted by the critical model $m_k$ must be handled by the optimal solution. A fundamental lower bound on the optimal solution is the maximum pressure any single model would exert if placed alone on an otherwise empty GPU, \ie, $KVPR_{OPT} \ge d_k/(C-w_k)$.
\end{itemize}
The final step involves integrating these insights to bound $KVPR_{max} = \frac{W_{before} + d_k}{S_{before} - w_k}$. 
Following Graham's proof~\cite{graham1969bounds}, we substitute these into the numerator and get $KVPR_{max}\le KVPR_{OPT}\cdot (1+\frac{C}{S_{g_{max}}-w_k})$.

\subsubsection{TP Support}
\label{ap:tp}
The model placement algorithm in Algorithm~\ref{alg:global} seamlessly integrates models utilizing Tensor Parallelism (TP). We conceptualize a TP model requiring $tp\_size$ GPUs as being composed of $tp\_size$ distinct parts.
For scheduling purposes, we create $tp\_size$ entries in the sorted model list for such a model, assigning each entry $\frac{1}{tp\_size}$ of the original weight and request rate. 
A beneficial property emerges from this decomposition: since these entries have identical $\frac{r_k}{s_k}$ values, they remain adjacent after sorting.
This adjacency increases the likelihood that, as the algorithm iterates, these parts are initially assigned to different GPUs due to rising KVPRs.
To ensure the distribution, if assigning a TP part to the GPU with the minimum KVPR would result in collocating it with another part of the same original model, we instead assign it to the GPU exhibiting the second-lowest KVPR. 
Through this decomposition strategy and modified assignment rule, our algorithm effectively considers and manages the placement of TP models alongside single-GPU models.

\subsection{Elastic Memory Overhead}
\label{sec:elastic-overhead}

\begin{figure}[t]
  \centering    \includegraphics[width=\linewidth]{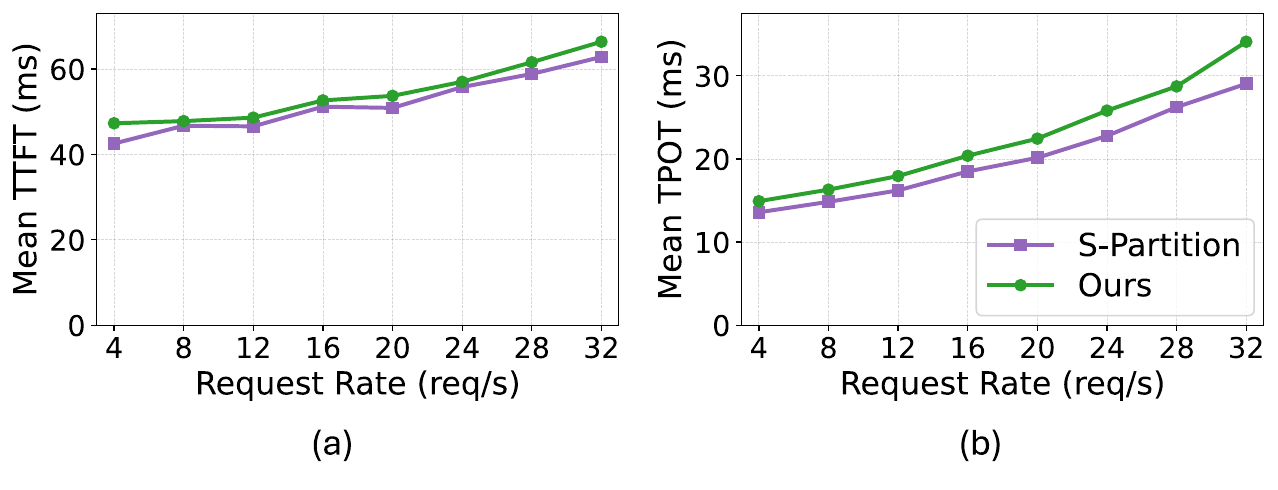}
    \vspace{-8mm}
    \caption{Mean latency comparison.
    }
    \label{fig:mean-latency-comparison}
    \vspace{-2mm}
\end{figure}

To stress-test the worst case, we evaluate the system under a constant request rate without idle periods, eliminating opportunities for memory coordination through either time- or space-sharing. In this setting, static partitioning provides the strongest baseline, as it divides memory according to the steady request rate. Figure~\ref{fig:mean-latency-comparison} presents a comparison between our system and static partitioning when serving two Llama-3.2-3B models on a single A100-40G GPU. In this experiment, the global scheduler does not alter placements and the GPU-local scheduler does not reorder requests, since all requests have equal priority. The only additional cost arises from the elastic memory manager, which dynamically maps and unmaps pages, unlike static partitioning that relies on pre-allocation. With optimizations such as buffer preallocation and contiguous layouts, this overhead remains modest: mean TTFT and TPOT increase by only 3–5\% as request rates scale, demonstrating the efficiency of our system even under worst-case conditions.

\subsection{Sensitivity Analysis}
\label{sec:sensitivity}

\begin{figure}[t]
  \centering    \includegraphics[width=\linewidth]{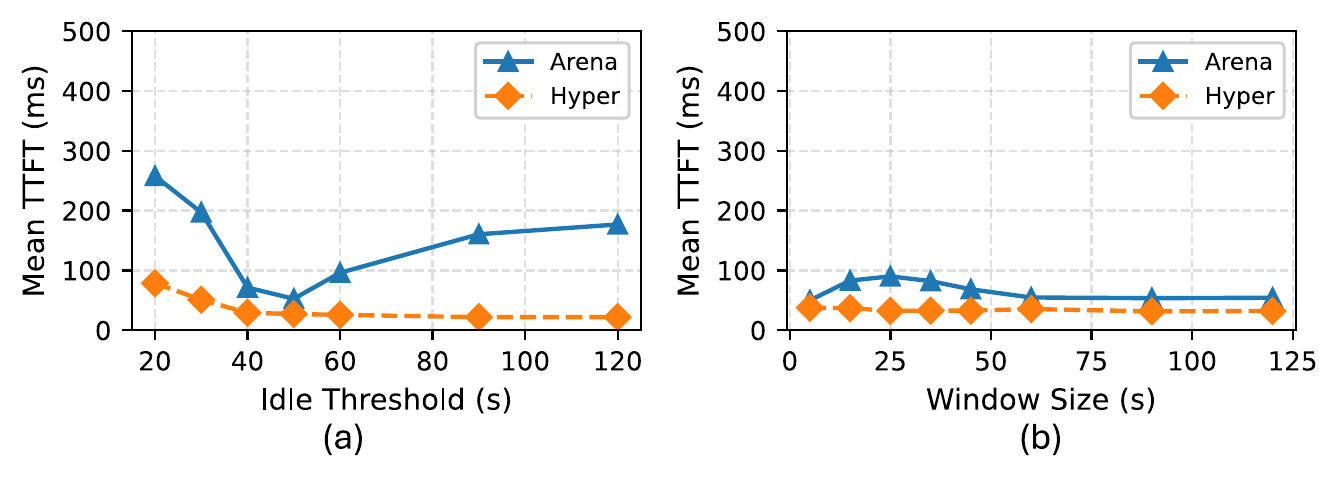}
    \vspace{-8mm}
    \caption{Sensitivity to hyperparameters.}
    \label{fig:sensitivity}
    \vspace{-2mm}
\end{figure}

We analyze the sensitivity of \tool to two key hyperparameters: the model eviction idle threshold and the load monitoring window size with Hyperbolic and Chatbot Arena traces. The results are shown in Figure~\ref{fig:sensitivity}.

\noindent \textbf{Idle Eviction Threshold.} 
Figure~\ref{fig:sensitivity}(a) reports the mean TTFT as we vary the idle threshold, which dictates how long a model must remain inactive before being evicted.
The results reveal a clear convex trade-off. 
When the threshold is too short (\eg, $<40$\,s), the system becomes overly aggressive, evicting models during short inter-arrival gaps. This leads to instability and thrashing, where the system must frequently pay the high penalty of model reactivation for subsequent requests, thereby inflating TTFT. 
Conversely, when the threshold is too long (\eg, $>80$s), idle models hoard GPU memory that could otherwise be harvested. This resource locking prevents the scheduler from placing new active models or expanding the KV cache for concurrent requests, resulting in resource starvation and increased queuing delays. 
Empirically, a threshold of approximately 45 seconds achieves the optimal balance between minimizing reactivation overhead and maximizing resource reclamation.

\noindent \textbf{Load Monitoring Window Size.}
Figure~\ref{fig:sensitivity}(b) examines the impact of the sliding window size used to calculate the moving average of token rates for the KVPR.
This parameter controls the sensitivity of the placement algorithm: a small window makes the scheduler sensitive to transient bursts, while a large window focuses on long-term trends.
The results demonstrate that \tool is generally robust to variations in window size. 
We observe that a window size of approximately 60 seconds provides a stable estimation of memory pressure, effectively smoothing out short-term noise while remaining responsive enough to shift resources during sustained workload changes.